\documentclass[aps
,pra
,amsmath
,eqsecnum
,amssymb
,amsfonts
,twocolumn
,letterpaper
,10pt
,tightenlines
,superscriptaddress
,nofootinbib
,eqsecnum
,longbibliography
,notitlepage
,floatfix
]{revtex4-2}

\usepackage{nag}
\usepackage{graphicx}
\usepackage{amsthm}
\usepackage{tabulary}
\usepackage{qcircuit}
\usepackage{mathtools}
\usepackage{bm}
\usepackage[table,xcdraw]{xcolor}
\usepackage{braket}
\usepackage{datetime}
\usepackage{stackrel}
\usepackage{stmaryrd}
\usepackage{dsfont}
\usepackage{qcircuit}
\usepackage[english]{babel}
\usepackage{units}

\usepackage{tikz}
\usetikzlibrary{backgrounds,decorations.pathreplacing}

\usepackage{chngcntr}
\counterwithout{equation}{section}

\usepackage[normalem]{ulem}
\usepackage{slashed}
\usepackage{soul}

\usepackage{xcolor}
\usepackage[
    colorlinks=true,
    urlcolor=blue,
    linkcolor=blue,
    citecolor=blue,
    filecolor=blue,
]{hyperref}
\usepackage[caption=false]{subfig}
\usepackage{amssymb}
\usepackage{amsmath}
\usepackage{amsthm}
\usepackage{braket}

\usepackage[OT2,T1]{fontenc}
\DeclareSymbolFont{cyrletters}{OT2}{wncyr}{m}{n}
\DeclareMathSymbol{\Sha}{\mathalpha}{cyrletters}{"58}

\usepackage{makeidx}
\makeindex

\usepackage{soul}
\usepackage{xargs}
\newcommandx{\cmnote}[2][1=]{\linespread{1.0}\todo[linecolor=red,backgroundcolor=red!25,bordercolor=red,#1]{#2}}

\let\underline\ul

\allowdisplaybreaks[4]

\DeclareMathOperator{\Tr}{Tr}

\DeclareMathOperator{\erf}{erf}

\renewcommand{\Im}{\operatorname{Im}}

\let\originalleft\left
\let\originalright\right
\renewcommand{\left}{\mathopen{}\mathclose\bgroup\originalleft}
\renewcommand{\right}{\aftergroup\egroup\originalright}

 \index{} \index{} \index{} \index{} \index{} \index{} \index{} \index{}
\makeatletter

\newcommand{\ringplus}{\mathbin{\text{\@ringplus}}}

\newcommand{\@ringplus}{%
  \ooalign{\hidewidth\raise1.3ex\hbox{\tiny$\circ$}\hidewidth\cr$\m@th+$\cr}%
}

\newcommand{\ringminus}{\mathbin{\text{\@ringminus}}}

\newcommand{\@ringminus}{%
  \ooalign{\hidewidth\raise0.9ex\hbox{\tiny$\circ$}\hidewidth\cr$\m@th-$\cr}%
}
\makeatother
 \index{} \index{} \index{} \index{} \index{} \index{} \index{} \index{}

\newcommand{\tp}[0]{\mathrm{T}}

\newcommand{\EPR}{\text{EPR}}
\newcommand{\bounceEPR}[2]{\op{A}_{#2}(#1)}
\newcommand{\bounceEPRgate}[2]{A_{#2}(#1)}

\newcommand{\GKP}{\text{GKP}} 
\newcommand{\GKPproj}{\op{\Pi}_\GKP}

\newcommand{\qunaught}{\varnothing}

\DeclareFontFamily{U}{wncy}{}
\DeclareFontShape{U}{wncy}{m}{n}{<->wncyr10}{}
\DeclareSymbolFont{mcy}{U}{wncy}{m}{n}
\DeclareMathSymbol{\Sh}{\mathord}{mcy}{"58}

\newcommand{\negspace}{\!}

\newcommand{\lsub}[2]{{\protect\vphantom{#1}}_{#2} \negspace {#1}}
\newcommand{\rsub}[2]{{#1} \negspace {\protect\vphantom{#1}}_{#2}}

\newcommand{\ketsub}[2]{\rsub {\ket{#1}} {#2}}
\newcommand{\brasub}[2]{\lsub {\bra{#1}} {#2}}

\newcommand{\pbra}[1]{\brasub{#1} p}

\newcommand{\pket}[1]{\ketsub{#1} p}
\newcommand{\qket}[1]{\ketsub{#1} q}

\newcommand{\outprod}[2]{\ket {#1}\!\bra {#2}}

\newcommand{\avg}[1]{\left\langle {#1} \right\rangle}

\newcommand{\abs}[1]{\left\lvert{#1}\right\rvert}

\newcommand{\reals}[0]{\mathbb{R}}

\newcommand{\op}[1]{\hat{#1}}

\newcommand{\opvec}[1]{\op{\vec{#1}}}

\newcommand{\id}[0]{I}

\newcommand{\mat}[1]{\bm{\mathrm{#1}}}

\renewcommand{\vec}[1]{\bm{#1}}

\newcommand{\controlled}[1]{\op{\mathrm{C}}_{#1}}
\newcommand{\CZ}[0]{\controlled Z}
\newcommand{\CX}[0]{\controlled X}

 \index{} \index{} \index{} \index{} \index{} \index{} \index{} \index{}
 \index{} \index{} \index{} \index{} \index{} \index{} \index{} \index{}

\newcommand{\red}{\color{red!90!black}}

\newcommand{\blk}{\color{black}}

 \index{} \index{} \index{} \index{} \index{} \index{} \index{} \index{}
 \index{} \index{} \index{} \index{} \index{} \index{} \index{} \index{}
\usepackage{graphicx}

\newcommand{\bsop}{\op{B} }

\usepackage{graphicx}
\usepackage{multirow}
\usepackage{hhline}

\xyoption{color}
\xyoption{line}

\newcommandx*\bsbal[3][1=black, 3=->]{\ar @[#1]@{#3} [#2,0] \qw}

\newcommandx*\varbs[4][1=black, 3=\theta,4=->]{\ar @[#1]@{#4}^{#3} [#2,0] \qw}

\newcommandx*\ctrlg[2]{\control \ar @{-}^{#1} [#2,0] \qw}
\newcommandx*\ctrlog[2]{\controlo \ar @{-}^{#1} [#2,0] \qw}

\newcommand{\EPRdr}{\ar @{-} [dr(0.5)]\qw }
\newcommand{\EPRur}{\ar @{-} [ur(0.5)]\qw }
\newcommand{\EPRul}{\ar @{-} [ul(0.5)] }
\newcommand{\EPRdl}{\ar @{-} [dl(0.5)] }

\makeatletter
 \newcommand{\xmapsfrom}[2][]{%
    \ext@arrow3095\leftarrowfill@{#1}{#2}\mapsfromchar
}
\makeatother

\newcommand{\QRL}{\text{QRL}}

\newcommand{\atantwo}{{\chi}}
\newcommand{\dB}{{\text{dB}}}

\newcommand{\IGKP}{\bar{I}}
\newcommand{\HGKP}{\bar{H}}
\newcommand{\PGKP}{\bar{P}}
\newcommand{\CZGKP}{\bar{\text{C}}_{Z}}

\newcommand{\SWAPGKP}{\overbar{\text{SWAP}}}

\newcommand{\overbar}[1]{\mkern 1.5mu\overline{\mkern-1.5mu#1\mkern-1.5mu}\mkern 1.5mu}

\begin{document}
\title{Streamlined quantum computing with macronode cluster states}
\newpage
\setcounter{page}{1}
\pagenumbering{arabic}

\author{Blayney W. Walshe}
\email{blayneyw@gmail.com}
\affiliation{Centre for Quantum Computation and Communication Technology, School of Science, RMIT University, Melbourne, VIC 3000, Australia}
\author{Rafael N. Alexander}
\affiliation{Centre for Quantum Computation and Communication Technology, School of Science, RMIT University, Melbourne, VIC 3000, Australia}
\author{Nicolas C. Menicucci}
\affiliation{Centre for Quantum Computation and Communication Technology, School of Science, RMIT University, Melbourne, VIC 3000, Australia}
\author{Ben Q. Baragiola}
\affiliation{Centre for Quantum Computation and Communication Technology, School of Science, RMIT University, Melbourne, VIC 3000, Australia}
\affiliation{Yukawa Institute for Theoretical Physics, Kyoto University, Kitashirakawa Oiwakecho, Sakyo-ku, Kyoto 606-8502, Japan}
\begin{abstract}
Continuous-variable cluster states allow for fault-tolerant measurement-based quantum computing when used in tandem with the Gottesman-Kitaev-Preskill (GKP) encoding of a qubit into a bosonic mode. For quad-rail-lattice macronode cluster states, whose construction is defined by a fixed, low-depth beam splitter network, we show that a Clifford gate and GKP error correction can be simultaneously implemented in a single teleportation step. We give explicit recipes to realize the Clifford generating set, and we calculate the logical gate-error rates given finite squeezing in the cluster-state and GKP resources.
We find that logical error rates of $10^{-2}$--$10^{-3}$, compatible with the thresholds of topological codes, can be achieved with squeezing of 11.9--13.7~dB. The protocol presented eliminates noise present in prior schemes and puts the required squeezing for fault tolerance in the range of current state-of-the-art optical experiments. Finally, we show how to produce distillable GKP magic states directly within the cluster state.
\end{abstract}
\maketitle

\section{Introduction}
\label{intro}
Continuous-variable (CV) cluster states are entangled resources for CV measurement-based quantum computation (MBQC)~\cite{Menicucci2006,Gu2009}. They are highly scalable, can be generated deterministically, and operate at room temperature---all of which make them an attractive substrate for quantum computing ~\cite{Yokoyama2013,Asavanant2019,Larsen2019,Raussendorf2001, Menicucci2011a,Chen2014,Yoshikawa2016}. 
CV cluster states were originally designed using single-mode squeezed states and CV controlled-$Z$ gates (in direct analogy to their qubit counterparts)~\cite{Menicucci2006}. The CV controlled-$Z$ gates require inline squeezing, which is experimentally difficult, however, and later work showed that this is unnecessary: CV cluster states can be made entirely with offline squeezing and passive linear optics~\cite{van2007building}, albeit with a design that makes scaling up to large sizes a challenge.
Later, it was realized that CV cluster states could be generated using an experimentally accessible set of resources: offline squeezing and constant-depth local linear optical circuits~\cite{Menicucci2008,menicucci2007ultracompact, flammia2009optical,Menicucci2011a, Wang2014, Alexander2016, Alexander2018, Wu2020, fukui2020temporal, zhu2021hypercubic, larsen2021architecture}. 

\begin{figure}[t!]
    \centering
    \includegraphics[width=0.9\columnwidth]{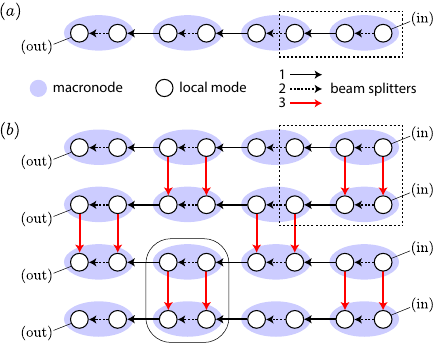} 
    \caption{\label{fig:macronodeWireGraph} Macronode cluster states for quantum computing. Each light purple oval designates a grouping of two local modes called a (two-mode) macronode. Arrows represent beam splitters between local modes, applied in the order $\{1,2,3\}$. (a) In the one-dimensional case, known as a macronode wire, macronodes are chained together using beam splitters. Macronode-local measurements teleport an input state along the macronode wire with gates applied at each macronode that depend on the measurement bases and the specific states in each wire. (b) An example of a two-dimensional quad-rail-lattice (QRL) construction.
    Macronode wires are periodically coupled to one another using additional beam splitters (vertical, red), and
    local measurements teleport multimode input states and facilitate two-mode gates.
    Previous work on the QRL interprets coupled two-mode macronodes as a four-mode macronode~\cite{alexander2016flexible};
    we circle one of these four-mode macronodes (solid outline) to highlight the defining property of the QRL: local four-mode coupling.
    Dashed boxes indicate the \emph{macronode gadgets} used to implement (a) single-qubit and (b) two-qubit Clifford gates on a GKP-encoded qubit. Their respective circuit diagrams can be found in Eq.~\eqref{circuit:macrnonodegadget} and Eq.~\eqref{twoModeCircuit_QRL}.
    }
\end{figure}

Since then, large-scale cluster states, which are all based on \emph{macronodes} (i.e.,~collections of multiple modes functioning as a single unit)~\cite{Menicucci2008,flammia2009optical,Menicucci2011a}, have been experimentally produced in the frequency~\cite{Chen2014} and temporal~\cite{Yokoyama2013,Asavanant2019,Larsen2019,Yoshikawa2016} domains. 
The macronode wire is a linear macronode cluster state (where each macronode has only two modes within it) used to implement single-mode Gaussian unitary operations in a measurement-based fashion.
It is constructed from a chain of two-mode squeezed states linked by 50:50 beam splitters~\cite{Menicucci2011a,Walshe2020} as shown in Fig.~\ref{fig:macronodeWireGraph}(a). 
Coupling together macronode wires using additional beam splitters produces higher-dimensional macronode cluster states that are useful for universal quantum computing~\cite{Menicucci2011a, Wang2014, alexander2016flexible, Alexander2016, Alexander2018, Wu2020, fukui2020temporal, zhu2021hypercubic, larsen2021architecture}. 
We focus on the quad-rail lattice (QRL)~\cite{Menicucci2011a,alexander2016flexible}, which is used to implement the two-mode unitaries required for universality. 
Although originally proposed as a macronode-based implementation of a two-dimensional square-lattice cluster state, the QRL construction~\cite{Menicucci2011a} does not require a specific graph topology---it only requires that four local modes are stitched together as in Fig.~\ref{fig:macronodeWireGraph}(b).
This can be used to realize a class of graphs that includes three-dimensional (3D) lattices, such as that in Ref.~\cite{Wu2020} and the Raussendorf--Harrington--Goyal~(RHG) lattice~\cite{Raussendorf2007,tzitrin2021fault}, which provide topological fault tolerance when used as a qubit cluster state. 

Any physical implementation of quantum circuits will involve some degree of noise, but CV MBQC also has to contend with \emph{intrinsic noise} that results from finite energy constraints~\cite{Gu2009, Alexander2014}. Thus, considerations of scalable quantum computing with CV cluster states require a method for addressing noise up front. 
Bosonic codes fill this role by encoding discrete-variable
quantum registers within the Hilbert space of one or more bosonic modes.
Reference~\cite{menicucci2014fault} proposed using the Gottesman-Kitaev-Preskill (GKP) encoding~\cite{GKP} to discretize the intrinsic Gaussian noise that arises in MBQC with CV cluster states. 
GKP error correction collapses CV noise (including intrinsic noise) into probabilistic qubit-level errors.
These errors will also need to be corrected, so they must be passed on to a higher level quantum error correcting code. Fault-tolerant quantum computation is possible given an effective qubit error rate below some threshold value (specific to the noise model, chosen higher level code, and decoder) \cite{menicucci2014fault,larsen2021architecture,Noh2021lowoverhead}.

Recent studies have married macronode cluster states with the GKP code~\cite{Walshe2020, Larsen2020noiseAnalysis} and found squeezing thresholds~\cite{larsen2021architecture, tzitrin2021fault, Noh2021lowoverhead} in reach of near-term technology. This provides a (non-optimized) target for experimental efforts into creating these resource states.
GKP error correction, which  mitigates intrinsic finite-squeezing noise, is possible out of the box, because the Gaussian unitary operations from MBQC with CV cluster states are all that are needed for the full set of single- and two-qubit GKP Cliffords~\cite{menicucci2014fault}.
However, current proposals that are based on local implementation of gates rely on compiling GKP Clifford gates and error correction across several teleportation steps~\cite{larsen2021architecture,Noh2021lowoverhead}, which is undesirable since each additional step adds finite-squeezing noise.

In this paper, we provide three critical advances. First, we simplify gate implementations by showing that all GKP Cliffords can be performed deterministically in a single teleportation step.
Second, we show that preparing the cluster-state modes in squeezed GKP states (called qunaught states~\cite{Walshe2020}), automatically implements GKP error correction during gate execution.
Together, these streamline quantum computing in the QRL construction by reducing the number of measurements per mode to just two per gate, which encompass both the gate itself and GKP error correction.
Third, we embed the scheme for making GKP magic states from Ref.~\cite{Baragiola2019} directly into the macronode setting, thereby eliminating any additional noise that would otherwise result from having to attach such states if they were prepared offline.

Using the first two advances, we calculate the logical-gate error rates for GKP Clifford implementations and find that they surpass previous best-case gate error rates~\cite{Larsen2020noiseAnalysis}. For the noisiest gate, the logical controlled-$Z$, we find that gate error rates of $10^{-2}$--$10^{-3}$ are achievable with 11.9--13.7~dB of squeezing in the resource states that comprise the cluster state. Since our advances use the minimum number of ancillae per gate, they outperform previous studies by at least $\sim 1.3$~dB~\cite{menicucci2014fault, Larsen2020noiseAnalysis, larsen2021architecture}.

We detail the assumptions made by our analysis in Section \ref{sec:squeezingthresholds}. Succinctly, we consider equivalent finite squeezing noise on all input states, and our GKP states are treated as if the Gaussian envelope width approaches infinity, which is justified in the limit of high squeezing~\cite{mensen2021,noh2020fault}. These assumptions are common in other literature (e.g.,~\cite{menicucci2014fault,bourassa2021blueprint,Larsen2020noiseAnalysis}), and we employ them for ease of comparison. We do not claim these gate error rates are actual fault-tolerance thresholds since such a claim would require a more sophisticated error model.

\section{Notation and conventions}\label{notation}
We first lay out notational conventions and important continuous-variable circuit identities used throughout this paper, most of which come from Ref.~\cite{Walshe2020}, where further details can be found.
\subsection{CV bases and unitary operators}
\label{subsec:CVunitaries}

Each CV mode has canonical position and momentum operators, $\op{q} = \frac{1}{\sqrt{2}}(\op{a} + \op{a}^\dagger)$ and $\op{p} = \frac{-i}{\sqrt{2}}(\op{a} - \op{a}^\dagger )$, obeying the canonical commutation relation $[\op{q}, \op{p}] = i$. This corresponds to an implicit choice of $\hbar = 1$, with measured vacuum variance equal to $\tfrac 1 2$ in both quadratures. Their eigenstates $\qket{s}$ and $\pket{t}$ satisfy $\op{q} \qket{s} = s \qket{s}$ and $\op{p} \pket{t} = t \pket{t}$.

The displacement operators
\begin{align}
    \op X (s) &\coloneqq e^{-i s \op p} = \op D(\tfrac {s} {\sqrt 2}),
\\
    \op Z (t) &\coloneqq e^{i t \op q} = \op D(\tfrac {i t} {\sqrt 2}),
\end{align}
displace by $+s$ in position and $+t$ in momentum, respectively: $\op X^\dag(s) \op q \op X(s) = \op q + s$, $\op Z^\dag(t) \op p \op Z(t) = \op p + t$, with $\op D(\alpha) = e^{\alpha \op a^\dag - \alpha^* \op a}$ being the ordinary quantum-optics displacement operator.

With the phase-delay operator 
    \begin{equation} \label{eq:phasedelay}
        \op R(\theta) \coloneqq e^{i \theta \op{a}^\dagger \op{a}}
        \, ,
    \end{equation}
we define a rotated momentum quadrature
    \begin{align} \label{eq:rotatedquadrature}
        \op{p}_{\theta} &\coloneqq \op{R}^\dagger(\theta) \op{p} \op{R}(\theta) =  
         \op p \cos \theta + \op q \sin \theta 
         \, .
    \end{align}
This quadrature's eigenstates, satisfying $ \op{p}_{\theta}  \ketsub{t}{p_{\theta}} = t \ketsub{t}{p_{\theta}}$, are given by
$\ketsub{t}{p_{\theta}} \coloneqq \op R^\dagger(\theta) \pket{t}$~\cite{Walshe2020}.

A special case of the phase delay operator is the Fourier transform operator, 
    \begin{equation} \label{eq:Fouriergate}
        \op{F} \coloneqq \op R(\tfrac{\pi}{2})
        \, ,
    \end{equation}
which rotates the canonical quadratures, $\op{F}^\dagger \op{q} \op{F} = -\op{p}$, and $\op{F}^\dagger \op{p} \op{F} = \op{q}$.
We describe measurements of a rotated quadrature $\op{p}_\theta$, realized via homodyne detection, as projections onto rotated eigenstates. 
In the circuit setting, these projections are described by the bra 
\begin{align}\label{rotmeasure}
    \brasub{m}{p_\theta}\coloneqq\pbra{m}\op {R}(\theta),
\end{align}
where $m$ is the measurement outcome. 

We use two additional single-mode Gaussian operators. First is the squeezing operator with squeezing factor~${\zeta \in \reals}$,
    \begin{align} 
         \op S(\zeta) & \coloneqq \op R(\Im \ln \zeta) e^{-\tfrac{i}{2} (\ln{\abs \zeta}) (\op q \op p-\op p \op q)} \label{eq:squeezinggate} \\
         &=\op R(\Im \ln \zeta) e^{\tfrac{1}{2} (\ln{\abs \zeta}) (\op a^{\dag 2} -\op a^2 )}
         \, ,
    \end{align}
where $\Im \ln \zeta = \pi$ if $\zeta < 0$ and zero otherwise. This appends a $\pi$ phase shift to the squeeze
if and only if $\zeta < 0$, keeping its symplectic Heisenberg action consistent for all real~$\zeta$ (see Ref.~\cite{Alexander2016}). 
Next is the momentum-shear operator with shear parameter~$\sigma$:
    \begin{align} 
         \op P(\sigma) &\coloneqq e^{\frac{i}{2}\sigma \op q^2}
         \, .\label{eq:sheargate} 
    \end{align}
    
The two-mode CV gates we focus on are the controlled-$Z$ gate and the balanced beam splitter. A controlled-$Z$ gate with real weight $g$, 
    \begin{align} \label{eq:CVCZ}
        \CZ(g) &\coloneqq e^{i g \op q \otimes \op q}
        \, ,
    \end{align}
is symmetric (invariant under swapping the inputs). 
The two-mode entangling gate for macronode cluster states is a balanced beam splitter. For modes $j$ and $k$, the beam splitter convention we use is
    \begin{align} \label{beam splitterdef}
        \bsop_{jk} &\coloneqq e^{-i \frac{\pi}{4}(\op{q}_j \otimes \op{p}_k - \op{p}_j \otimes \op{q}_k )}\\
        &= e^{-\tfrac{\pi}{2}(\op a_j \otimes \op a_k^\dag-\op a_j^\dag \otimes \op a_k)}
        \, .
    \end{align}
Note that Hermitian conjugation is equivalent to exchanging the inputs: $\op B_{jk}^\dag = \op B_{kj}$.

\subsection{Quantum circuits and right-to-left convention}

Following Ref.~\cite{Walshe2020}, the circuits in this paper flow \emph{from right to left}, with input states specified by kets on the right-hand side of the circuit and projective measurements (including the outcome) specified by bras on the left-hand side. With this convention, which is sometimes called the Kitaev convention, each circuit maps simply to Dirac notation. This means gates merge together without reversing their order, and projective measurements can be straightforwardly represented as bras. For instance,
\label{eq:RtoLcircuit}
\begin{align}
    \quad\;\;
    \Qcircuit @C=0.6em @R=1em
    {
	 &\lstick {\brasub{m}{b}\!} & \gate B & \gate A & \rstick{\!\ket{\psi}} \qw
	 }
	\qquad=\quad\;\;\;
    \Qcircuit @C=0.6em @R=1em
    {
	& \lstick {\brasub{m}{b}\!} & \gate {B A} & \rstick{\!\ket{\psi}} \qw
	 }
	\qquad=\,
    \brasub{m}{b} \op B \op A \ket \psi
	\raisebox{-0.5ex}{,}
\end{align}
where $\ket \psi$ is the input state, $\brasub{m}{b}$ indicates a measurement in basis~$b$ with outcome~$m$, and the result of the circuit is an amplitude for that outcome. Circuits for which only some of the systems are measured produce Kraus operators associated with that outcome under a similarly straightforward mapping. The notation for other circuit elements is standard and without modification, except for the understanding that time flows right to left.

One circuit element of particular importance, whose novel notation was first introduced in Ref.~\cite{Walshe2020}, is that of the beam splitter, Eq.~\eqref{beam splitterdef}. We represent this as a vertical arrow pointing from the wire for mode~$j$ to that for mode~$k$:
\begin{equation}\label{BScircuit}
\raisebox{-1em}{$\bsop_{jk} =\quad {\scriptsize\text{(out)}}~~$}
\Qcircuit @C=1.25em @R=2.1em {
	 &
	 \bsbal{1}
	 &
	 \rstick{j}
	 \qw  \\
	 &
	 \qw
	 &
	 \rstick{k}
	 \qw
		}
	\raisebox{-1em}{\quad\;\; {\scriptsize\text{(in)}} \quad.}
\end{equation}
Since $\op B_{jk}^\dag = \op B_{kj}$, taking the Hermitian conjugate reverses the direction of the arrow.

\subsection{Gottesman-Kitaev-Preskill code}

The ideal, square-lattice GKP computational basis states, indexed by $j \in \{0,1\}$, are described by periodic wavefunctions in position and momentum, respectively given by~\cite{GKP}
\begin{align}\label{GKP}
    \ket{j_\GKP} &\coloneqq 
		\int ds\, \Sha_{2\sqrt\pi}( s-j\sqrt{\pi} ) \qket s\\
        &= \int dt\, e^{ij\sqrt{\pi}t}\Sha_{\sqrt{\pi}}(t)\pket{t},
\end{align}
where a Dirac comb of period $T$ is defined as~\cite{mensen2021}
	\begin{equation}
		\Sha_{ T }(x) \coloneqq \sqrt{T} \sum_{n=-\infty}^\infty \delta(x - n T)
	.
	\end{equation}
Together, the basis states span a two-dimensional subspace of a CV mode's Hilbert space that is described by the projector 
\begin{align} \label{eq:GKPproj}
    \GKPproj \coloneqq &\outprod{0_\GKP}{0_\GKP} + \outprod{1_\GKP}{1_\GKP} 
         \, .
\end{align}

A distinguishing feature of GKP codes is that the Clifford group can be implemented with Gaussian unitary operations on the CV mode. For the square-lattice GKP code considered here, the correspondence between Gaussian CV unitaries and their action as Clifford gates in the square-lattice GKP encoding is
    \begin{align} \label{eq:gateconnections}
        \underbrace{\big\{ \op{I}, \op{F}, \op{P}(\pm 1), \CZ(\pm 1)\big\}}_{\text{CV unitaries}} 
        \longmapsto 
        \underbrace{\big\{ \IGKP, \HGKP, \PGKP, \CZGKP \big\}}_{\text{GKP Cliffords}}
        ,
    \end{align}
respectively. 
The CV unitaries were introduced in Sec.~\ref{subsec:CVunitaries}, and the GKP Clifford gates use standard notation for qubit gates, with $\PGKP$ indicating the phase gate ($\tfrac \pi 2$ rotation about the $Z$ axis of the Bloch sphere). 
Throughout this paper, CV unitaries are indicated by hats and logical GKP gates are indicated by overbars.
Many CV unitaries can perform the same logical gate on a square-lattice GKP state---for example, $\op{F}^\dagger$ also implements $\HGKP$; see Ref.~\cite{GKP} for further details.
\section{Quantum computing with quad-rail-lattice cluster states and the GKP code}
\label{sec:QCmacronode}

The key to using CV cluster states for computing with the GKP code is determining the measurement bases that implement GKP Clifford gates.
For a slightly different macronode cluster state, Larsen \emph{et al.} recently proposed a set of gates (including two-qubit Clifford gates)~\cite{larsen2021architecture} that requires at most two steps (teleportation through two macronodes) and relies on variable-transmission beam splitters for error correction.

In this paper, we give an improved protocol that provides three advances. First, the 
full generating set of GKP Clifford operations can be performed in a single measurement step. That is, all single-qubit Cliffords are executed during teleportation through a single macronode, and the two-qubit Cliffords are executed through two entangled macronodes. This more efficient use of the macronode cluster state reduces the amount of finite-squeezing noise per gate.  Second, GKP error correction is performed in parallel with each logical gate by teleporting through an encoded GKP Bell pair. This introduces less noise than an ancilla-coupled approach and leads to the improved gate error rates in Sec.~\ref{sec:squeezingthresholds}.
Third, our protocol does not require variable-transmission beam splitters; rather, the beam splitter network is fixed, which simplifies experimental implementation.
\subsection{Single-mode gates}
We describe the above concepts in more detail using the essential primitive element in a macronode wire---the macronode teleportation gadget, indicated by a dashed black box in Fig.~\ref{fig:macronodeWireGraph}(a). Each macronode gadget consists of three modes. Measuring the first two teleports the input state to the output mode with a Kraus operator applied. The circuit identity for the macronode gadget, derived in Ref.~\cite{Walshe2020}, is 
  \begin{align}\label{circuit:macrnonodegadget}
  \resizebox{\columnwidth}{!}
  {
    \phantom{$\scriptsize ~p_{\theta_a}$}\Qcircuit @C=1.75em @R=2em 
    {
	&\lstick{\brasub{m_a}{p_{\theta_a}}}   & \bsbal{1} & \qw & \rstick{\text{(in)}} \qw[-1] &  \\
	&\lstick{\brasub{m_b}{p_{\theta_b}}}  & \qw   &  \bsbal{1} & \rstick{\ket{\psi}} \qw \\
	&\lstick{\text{(out)}}	& \qw  &   \qw  & \rstick{ \ket{\phi} } \qw
		}\, 
	\raisebox{-2.2em}{~=} 
    \Qcircuit @C=1em @R=1em {
    &\ar @{-} [dl(0.5)] &\gate{ D(\mu_{a,b}) } &\gate{\frac{1}{\sqrt{\pi}} V(\theta_a ,\theta_b)}      &\rstick{\text{(in)}} \qw[-1] &     \\
    &\ar @{-} [ul(0.5)] &\qw             &\qw                                                    &\qw[-1] \ar @{-} [dr(0.5)] &&   \\
	&&\lstick{\text{(out)}}      
    &\gate{\frac{1}{\sqrt{\pi}}\bounceEPRgate{\psi,\phi}{}} &\qw[-1] \ar @{-} [ur(0.5)] &&
		}
	}
	\!\!\raisebox{-2em}{,}
    \end{align}
where the specific elements in the right hand side are defined and discussed below. 
We call this circuit the \emph{single-mode macronode gadget} since it takes a single mode as input and is implemented at a single (two-mode) macronode. In fact, this circuit is identical to the standard CV teleportation circuit~\cite{Furusawa1998}. Note that the term `single-mode' refers to the number of inputs, with the size of the macronode being double that since two measurements are required to teleport a single mode.

The gate $\op V(\theta_a,\theta_b)$ is a Gaussian unitary determined by the
homodyne measurement angles $\theta_a$ and $\theta_b$. 
We give the standard form of this unitary (up to an overall phase)~\cite{Alexander2014, Walshe2020} along with a decomposition that will be useful when working with the GKP code:
    \begin{align}\label{vGateOrig}
    \op V(\theta_a,\theta_b)&=\op R(\theta_+-\tfrac{\pi}{2})\op S(\tan{\theta_-})\op R(\theta_+)\\
 &= \op R(\theta_a-\tfrac{\pi}{2}) \op P[2\cot (2 \theta_-)] \op R(\theta_a-\tfrac{\pi}{2}) \label{vGateqShear} 
 \, , 
\end{align}
where the operators on the right are defined in Sec.~\ref{subsec:CVunitaries}, and
    \begin{align}
        \theta_\pm &\coloneqq \frac{\theta_a \pm \theta_b}{2}
        .
    \end{align}
Equation~\eqref{vGateqShear}
arises from the Bloch-Messiah decomposition~\cite{braunstein2005squeezing} of the squeezing operation (up to an overall phase), 
    \begin{align} \label{eq:localsqzdecomp}
    \op S(\tan \theta_-) 
    &= \op R(\theta_-) \op P[2 \cot (2\theta_-)] \op R(\theta_- - \tfrac \pi 2)
    \, ,
    \end{align}
which can be verified directly using the symplectic representation of the Heisenberg action of these operators~\cite{Alexander2014}.

The amplitude of the displacement $\op {D}(\mu_{a,b})$ depends on the measurement outcomes $m_a$ and $m_b$ and on the chosen measurement bases:
    \begin{equation} \label{eq:mu}
        \mu_{a,b}
        \coloneqq \frac{ - m_a e^{i \theta_b} - m_b e^{i \theta_a } }{ \sin (2 \theta_-)}
        \, .
    \end{equation}
At each macronode, this displacement is known and can be corrected. For this reason, we frequently set $m_a = m_b = 0$ throughout this paper so that $\mu = 0$. Interested readers can consult Ref.~\cite{Walshe2020} for more details. 

Finally, the local states $\ket{\psi}$ and $\ket{\phi}$ at the input
determine the gate $\op {A}(\psi,\phi)$, which is, in general, not unitary. The precise definition of this gate and more details about it can be found in Ref.~\cite{Walshe2020}. When its particular form is required, we will give it explicitly for that special case.

The circuit in Eq.~\eqref{circuit:macrnonodegadget} describes the Kraus operator 
    \begin{equation} \label{genkraus}
        \op{K}(m_a, m_b) = \frac{1}{\pi} \bounceEPR{\psi, \phi}{} \op D(\mu_{a,b}) \op{V}(\theta_a,\theta_b) 
        \, ,
    \end{equation}
which gives the evolution of an input state $\op{\rho}_\text{in}$ as it is teleported from the top mode to the bottom mode:
    \begin{align} \label{eq:Krausmap}
        \op{\rho}_\text{out} = \frac{1}{\Pr(m_a, m_b)} \op{K}(m_a, m_b) \op{\rho}_\text{in} \op{K}^\dagger(m_a, m_b)
        \, .
    \end{align}
Since the Kraus operator is not unitary, the output state is renormalized by the probability density of the outcomes, 
$\Pr(m_a, m_b) = \Tr [\op{K}^\dag(m_a, m_b) \op K(m_a, m_b) \op{\rho}_\text{in} ]$. 

The macronode gadget, Eq.~\eqref{circuit:macrnonodegadget}, allows us to implement a deterministic single-mode Gaussian unitary $\op {V}(\theta_a,\theta_b)$, Eq.~\eqref{vGateOrig}, through a choice of homodyne measurement bases. (It is deterministic because the displacement is known and can be corrected at the end of the gadget or accounted for in later steps of the computation.)
Two consecutive $\op {V}(\theta_a,\theta_b)$ gates can enact any single-mode Gaussian operation~\cite{Alexander2014}. Such generality is not required for the GKP encoding, however. In fact, the minimal set of single-qubit GKP Clifford gates in Eq.~\eqref{eq:gateconnections} can be performed in a \emph{single step}---i.e.,~a single macronode gadget---using the 
measurement bases presented in Table~\ref{tab:twoModeMeasurementAngles}. 
This is a distinctive feature of GKP Clifford gates. Single-step operation allows for lower gate error rates than previously reported (discussed further in Sec.~\ref{sec:squeezingthresholds}).
 
 \begin{table}[t]
\def\arraystretch{1.5}
\begin{tabular}{|c|c|c|c|}
\hline
$\{\theta_a,\theta_b\}$ & $\op V(\vec \theta)$ & Logical Gate\\
\hline
$\{\tfrac{\pi}{2},0\}$ & $\op I$ & $\Bar{I}$\\
$\{\tfrac{3\pi}{4},\tfrac{\pi}{4}\}$ & $\op F$ & $\Bar{H}$\\
$\{\tfrac{\pi}{2},\tfrac{\pi}{2}\mp \atantwo \}$ & $\op P(\pm 1)$ & $\Bar{P}$\\
\hhline{|===|}
  $\{\theta_a,\theta_b,\theta_c,\theta_d\}$
 & $\op {V}^{(2)}(\vec{\theta})$ & Logical Gate \\
  \hline
  $\{\tfrac{\pi}{2},\tfrac{\pi}{2}\pm \atantwo,\tfrac{\pi}{2},\tfrac{\pi}{2} \mp \atantwo\}$& $\CZ(\pm 1)$ &$\CZGKP$ \\
   $\{0,\tfrac \pi 2,\tfrac \pi 2,0 \}$& SWAP & $\SWAPGKP$ \\
   $\{\tfrac{\pi}{2},0, \tfrac{\pi}{2},0\}$&$ \op I \otimes \op I$ &$\bar{I} \otimes \bar{I} $\\
   $\{\frac{3 \pi}{4},\tfrac{\pi}{4},\frac{3 \pi}{4},\tfrac{\pi}{4}\}$& $\op F\otimes \op F$ & $\bar{H} \otimes \bar{H}$\\
   $\{\frac{\pi}{2},\tfrac{\pi}{2}\mp \atantwo,\frac{ \pi}{2},\tfrac{\pi}{2}\mp \atantwo\}$& $\op P(\pm 1)\otimes \op P(\pm 1)$ & $\bar{P} \otimes \bar{P}$ \\
   \hline
\end{tabular}
\caption{
Measurement bases and the resulting GKP Clifford gates for the QRL macronode cluster state. 
The upper set (above the double horizontal line) are single-mode CV gates realized with the single-mode macronode gadget in Eq.~\eqref{circuit:macrnonodegadget}. These gates implement single-qubit GKP Cliffords.
The lower set (below the double horizontal line) are two-mode CV gates realized via the two-mode macronode gadget in Eq.~\eqref{twoModeCircuit_QRL}. This set includes an entangling GKP Clifford gate, the SWAP gate, and several  pairs of identical single-mode gates.
The constant angle $\atantwo$ is defined as $\atantwo \coloneqq \arctan{2 }\approx 1.1071~\mathrm{rad} \approx 63.435^\circ$.
}
\label{tab:twoModeMeasurementAngles}
\end{table}

\subsection{Two-mode gates}
Thus far, we have considered only single-mode gates implemented by measurements on a macronode wire. To complete the Clifford group, we need to implement an entangling gate between encoded GKP qubits. 
A suitable resource for this purpose is created by coupling macronode wires together via additional beam splitters into a two-dimensional lattice. 
There are various ways to do this~\cite{Larsen2019,Asavanant2019,Alexander2016,Menicucci2011a}, distinguished by (among other things) the number of modes per macronode in the final state. Here, we focus on an architecture with four modes per macronode, called the QRL~\cite{Menicucci2011a,alexander2016flexible,Alexander2017},
which was previously found to have favorable noise properties compared to other lattices~\cite{Larsen2020noiseAnalysis}.
The QRL construction depends only on each macronode comprising exactly four modes. Thus, when we refer to ``the QRL,'' we mean the particular method of stitching together a four-node macronode~\cite{Menicucci2011a,alexander2016flexible}, with the understanding that this can be applied to any graph of degree 4---i.e.,~a graph in which every node is connected to exactly four neighbors. This is an important class of graphs that includes, among others, the RHG lattice~\cite{Raussendorf2007} for which a QRL construction has been proposed and a squeezing threshold found~\cite{Tzitrin2020}.
In any QRL-based architecture, groups of four modes are coupled using a four-splitter~\cite{alexander2016flexible}, which is typically implemented using four beam splitters. 
One can equally well consider two of these beam splitters to be first creating macronode wires, as shown by the dashed arrows in Fig.~\ref{fig:macronodeWireGraph}(a) and also in (b).
Macronode wires are then stitched together at a four-mode macronode using the remaining two beam splitters, as shown by the red (vertical) arrows in Fig.~\ref{fig:macronodeWireGraph}(b), with an example of a four-mode macronode circled (with a solid border) in that figure.

Each four-mode macronode powers a two-mode gadget capable of implementing either two single-mode gates or one two-mode gate~\cite{alexander2016flexible}. We illustrate both actions together using a single quantum circuit, which corresponds directly to the dashed box of Fig.~\ref{fig:macronodeWireGraph}(b): 
\begin{equation}\label{twoModeCircuit_QRL}
    \begin{split}
\Qcircuit @C=1.4em @R=2em {
		&&& \lstick{\brasub{m_a}{p_{\theta_a}}}  &  \qw & \qw &  \bsbal{3}[-->] & \bsbal{1}[.>] & \qw &\qw & \rstick{\text{(in)}} \\
		&&& \lstick{\brasub{m_b}{p_{\theta_b}}}  & \qw  &\bsbal{3}[-->] & \qw &  \qw & \bsbal{1} & \qw & \rstick{\ket{\psi}}& \\
		 \lstick{\text{(out)}} & \qw& \qw& \qw & \qw& \qw &\qw & \qw & \qw & \qw& \rstick{\ket{\phi}}& \\
		&&& \lstick{\brasub{m_c}{p_{\theta_c}}} & \qw & \qw & \qw & \bsbal{1}[.>] & \qw & \qw&\rstick{\text{(in)}} \\
		&&& \lstick{\brasub{m_d}{p_{\theta_d}}}  &\qw& \qw & \qw & \qw & \bsbal{1}&\qw & \rstick{\ket{\psi'}}& \\
		 \lstick{\text{(out)}} & \qw& \qw&\qw & \qw & \qw &\qw & \qw & \qw&\qw & \rstick{\ket{\phi'}}& \
}
\end{split}
\quad ,
\end{equation}
where the beam splitters are all the same 50:50 beam splitter with dashed arrows corresponding to the red, vertical ones in Fig.~\ref{fig:macronodeWireGraph}. We call this circuit the \emph{two-mode macronode gadget} since it takes two modes as inputs, produces outputs over two modes, and is implemented at a single (four-mode) macronode. It is the two-mode generalization of the single-mode macronode gadget shown in Eq.~\eqref{circuit:macrnonodegadget}.

This macronode gadget generates a two-mode quantum operation (which could be separable or entangling) between the input modes as they are jointly teleported to the output modes. Just as for the single-mode macronode gadget, Eq.~\eqref{circuit:macrnonodegadget}, the two-mode quantum operation that gets implemented depends on the quadratures being measured, the measurement outcomes $\vec{m}$, and the ancilla states that comprise the two-mode gadget.

The two-mode Kraus operator
    \begin{equation}\label{eq:twomodekraus_QRL}
        \op K^{(2)}(\vec m)= \frac{1}{\pi^2} \big[ \op {A}_1(\psi,\phi) \otimes \op {A}_2(\psi',\phi')\big] \op{D}_\QRL^{(2)}(\vec{m} ) \op V_\QRL^{(2)}(\vec \theta)
    \end{equation}
has three parts. First, each macronode contributes a teleported gate $\op{A}_j(\psi,\phi)$ with the subscript indicating the output mode on which the gate acts. Second, each contributes an outcome-dependent displacement, and the two displacements are mixed by the beam splitters into
    \begin{align}\label{eq:QRLdisplacements}
        \op{D}_\QRL^{(2)}(\vec{m} ) \coloneqq \op{D}_1(\mu_+) \otimes \op{D}_2(\mu_-)
        \, ,
    \end{align}
with $\mu_\pm = \tfrac{\mu_{c,d}\pm\mu_{a,b}}{\sqrt{2}}$.
Finally, the quadrature measurement bases $\vec \theta = ( \theta_a, \theta_b, \theta_c, \theta_d )$ implement the deterministic two-mode Gaussian unitary (up to a global phase), 
    \begin{align} \label{VtwoMode_QRL}
        \op{V}_\QRL^{(2)}(\vec \theta) \coloneqq \bsop_{21} [\op{V}_1(\theta_a, \theta_b) \otimes \op{V}_2(\theta_c, \theta_d)] \bsop_{12}
        \, ,
    \end{align}
which is represented by the circuit,
\begin{equation}\label{VtwoMode_QRLcircuit}
\Qcircuit @C=1.4em @R=1.4em {
		&\qw&\gate{V_1(\theta_a, \theta_b)}  &  \bsbal{1} &\qw \\
		&\bsbal{-1}& \gate{ V_2(\theta_c, \theta_d) } & \qw  &\qw \\
}
\, \raisebox{-1.5em}{.}
\end{equation}
A derivation of this gate is given in Appendix~\ref{TwoModeV}.

Choosing various measurement angles allows us to realize various two-mode Gaussian unitaries. 
Most important for GKP quantum computing is a two-mode Clifford gate, which can be implemented by a CV controlled-$Z$ gate $\CZ(\pm1)$; see Eq.~\eqref{eq:gateconnections}. 
To find the measurement angles that realize this gate, 
we decompose a CV controlled-$Z$ gate of weight $g$ as
    \begin{align}
        \CZ(g) =\bsop_{21} [\op{P}(-g) \otimes \op{P}(g)] \bsop_{12}
        \, ,
    \end{align}
equivalently described by the circuit identity
\begin{equation}\label{}
\begin{split}
\raisebox{0.6em}{
\Qcircuit @C=1.4em @R=2.6em {
		&&\ctrlg{g}{1}&\qw \\
		&& \ctrl{-1}&\qw \\
}
}
\, 
\end{split}
=
    \begin{split}
\Qcircuit @C=1.4em @R=1.2em {
		& \qw           &\gate{P(-g)}    &  \bsbal{1} &\qw \\
		&\bsbal{-1}& \gate{P(g)} & \qw                   &\qw \\
}
\end{split}
\; ,
\end{equation}
with the left-hand side being the circuit for $\CZ(g)$.\footnote{A related decomposition in terms of local squeezing between beam splitters is given in Ref.~\cite{Kalajdzievski2021}. That decomposition is related to ours through Eq.~\eqref{eq:localsqzdecomp}.}
This convenient decomposition allows us to implement the gate using two single-mode shears of equal magnitude and opposite sign.
When $g=\pm 1$, the gate acts as the two-qubit GKP Clifford $\CZGKP$ gate we desire; measurement angles that produce this gate are given in Table~\ref{tab:twoModeMeasurementAngles}. We note that there are many other sets of measurement angles that implement Clifford equivalent entangling gates, which we will detail in future work.

In Table~\ref{tab:twoModeMeasurementAngles} we also include the CV SWAP gate that exchanges states across the two modes, $\text{SWAP} \ket{\psi} \otimes \ket{\phi} = \ket{\phi} \otimes \ket{\psi}$. This includes the case when those states are GKP encoded.

We also  review  a method of disentangling the two-mode gate~\cite{alexander2016flexible}, allowing more versatile use of the QRL. When the single-mode Gaussian unitaries $\op{V}$ in Eq.~\eqref{VtwoMode_QRLcircuit} are identical, they commute with the beam splitters, which then cancel:
\begin{equation}\label{}
    \begin{split}
\Qcircuit @C=1.4em @R=1.5em {
		&\qw&\gate{V}  &  \bsbal{1} &\qw \\
		&\bsbal{-1}& \gate{V} & \qw  &\qw \\
}
\, 
\end{split}
=
 \begin{split}
\Qcircuit @C=1.4em @R=1.5em {
		&\qw  &  \bsbal{1}&\gate{V} &\qw \\
		&\bsbal{-1} & \qw& \gate{V}  &\qw \\
}
\, 
\end{split}
=
 \begin{split}
\Qcircuit @C=1.4em @R=1.5em {
		&\gate{V} &\qw \\
		& \gate{V}  &\qw \\
}
\end{split}
\; .
\end{equation}
Thus, one can implement two identical single-mode GKP gates  simultaneously on both modes with the two-mode macronode gadget. Several useful examples are given in Table~\ref{tab:twoModeMeasurementAngles}.

\subsection{Teleporation-based GKP error correction}

In addition to the measurement-basis-dependent gates discussed above, the macronode gadget in Eq.~\eqref{circuit:macrnonodegadget} also applies a quantum operation $\op{A}(\psi,\phi)$ that depends on the states $\ket{\psi}$ and $\ket{\phi}$~\cite{Walshe2020}. For an ideal macronode-based CV cluster state (with no qubit encoding),  $\ket{\psi} = \pket{0}$ and $\ket{\phi} = \qket{0}$, which is depicted as
    \begin{equation}
    \begin{split}\label{KnillCartoonQP}
    \centering
    \includegraphics[width=.65\columnwidth]{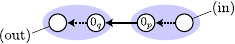}
    \end{split}\, ,
    \end{equation}
in the schematics of Fig.~\ref{fig:macronodeWireGraph}.
This choice generates a maximally entangled Einstein-Podolsky-Rosen (EPR) pair across two neighboring macronodes~\cite{Walshe2020}, giving the trivial operation
    \begin{align} \label{eq:teleportedidentity}
        \op{A}(0_p,0_q) = \op{I}
        \, 
    \end{align}
that forms the backbone of standard CV teleportation.

GKP error correction can be performed automatically when the states in a macronode gadget are themselves GKP states~\cite{Walshe2020}. However, since beam splitters introduce additional squeezing that modifies the spacing of a periodic wavefunction, the appropriate states are not square-lattice $\ket{+_\GKP}$ states. Rather, they are Fourier-invariant \emph{qunaught} states~\cite{Walshe2020} (also called sensor states~\cite{Duivenvoorden2017}), 
with wavefunction period $T=\sqrt{2\pi}$ in both position and momentum,
    \begin{equation} \label{qunaught}
        \ket{\qunaught} := \int ds \; \Sha_{\sqrt{2\pi}}(s)\qket{s}
        =\int dt \; \Sha_{\sqrt{2\pi}}(t)\pket{t} 
        \, ,
    \end{equation}
with the empty-set symbol~$\qunaught$ and `naught' in the name indicating that the state carries no quantum information~\cite{Walshe2020}.
Nevertheless,  combining two qunaught states on a beam splitter produces  an encoded Bell pair of square-lattice GKP qubits~\cite{Walshe2020}, 
      \begin{equation} \label{eq:GKPBellPair}
         \bsop_{12}\ket{\qunaught} \otimes \ket{\qunaught}
    = \tfrac{1}{\sqrt{2}}\big( \ket{0_\GKP} \otimes \ket{0_\GKP}  + \ket{1_\GKP} \otimes \ket{1_\GKP} \big)
    \, .
     \end{equation} 
     
A GKP Bell pair across neighboring macronodes is produced by preparing both $\ket{\phi}$ and $\ket{\psi}$ in the macronode gadget, Eq.~\eqref{circuit:macrnonodegadget}, in qunaught states:
    \begin{equation}
    \begin{split}\label{KnillCartoonQunaught}
    \centering
    \includegraphics[width=.65\columnwidth]{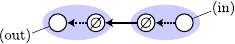}
    \end{split}\, .
    \end{equation}
Teleportation through an encoded Bell pair is the foundation for Knill-style GKP error correction~\cite{Knill2005}, and indeed this choice yields the quantum operation
    \begin{align} \label{eq:teleportedGKPproj}
        \op{A}(\qunaught,\qunaught) = \sqrt{ \frac{\pi}{ 2 } } \GKPproj
        \, ,
    \end{align}
with $\GKPproj$ being the projector onto the square-lattice GKP subspace [Eq.~\eqref{eq:GKPproj}]. From the Kraus operator in Eq.~\eqref{genkraus}, we see that this allows a GKP Clifford gate followed by GKP error correction to be performed in the same step (\emph{i.e.},~teleportation through a single macronode). 
The two-mode macronode gadget works identically, since the quantum operations $\op{A}(\psi,\phi)$ are local to each of the output modes as shown in the two-mode Kraus operator,  Eq.~\eqref{eq:twomodekraus_QRL}. Preparing all four ancilla states in the gadget in $\ket{\qunaught}$ implements the two-mode Gaussian gate determined by $\vec\theta$ and $\vec m$ followed by GKP error correction on each output mode.

\subsection{Magic states}
\label{subsec:magicstates}

Extending the above described operations to a universal gate set requires a logical non-Clifford element. Given access to high-fidelity Clifford gates, a universal set of operations is attainable by probabilistically distilling a high-quality magic state from multiple noisier copies via Clifford operations~\cite{bravyi2005universal}. 
Two commonly considered magic states are the $\ket{+T}$ state (stabilized under a Clifford gate that permutes the positive Pauli axes), 
\begin{align}
    \ket{+T} 
     &= \biggl(\frac{\sqrt{3}+3}{6} \biggr)^{\frac{1}{2}} \ket{0} + e^{i \pi/4} \biggl( \frac{2-\sqrt{3}}{6} \biggr)^{\frac{1}{4}} \ket{1}
     \, ,
    \end{align}
and the $\ket{+H}$ state (stabilized under the Hadamard gate),
\footnote{Both of these magic states and their Clifford equivalents can be used to deterministically teleport non-Clifford gates (given Clifford resources). Unfortunately, they go by different names in the literature as do the gates they teleport. We follow the conventions in Ref.~\cite{bravyi2005universal}.}
    \begin{align}
    \ket{+H} 
    &= \cos\tfrac{\pi}{8}\ket{0} + \sin\tfrac{\pi}{8}\ket{1}
    \, .
\end{align}
These states are distillable with ideal Clifford operations from noisy copies with fidelities no worse than 0.853~\cite{reichardt2005quantum} and 0.8273~\cite{jochym2013robustness}, respectively. 
Distillation of magic states with imperfect (but high-fidelity) Clifford gates is possible but requires copies with higher fidelities. 

An experimentally convenient method for probabilistically generating magic states of a desired fidelity in the GKP code was introduced in Ref.~\cite{Baragiola2019}. Performing GKP error correction on the vacuum state (or a low-photon-number thermal state)
produces a heralded, distillable GKP magic state with high probability.\footnote{The analysis in Ref.~\cite{Baragiola2019} focuses on vacuum and thermal states, but many other Gaussian states of high purity will also yield heralded, distillable magic states. Notable exceptions are position- or momentum-squeezed vacuum states, which are often used in combination with GKP states to produce hybrid cluster states for use in fault-tolerant architectures~\cite{menicucci2014fault, bourassa2021blueprint, larsen2021architecture}. These Gaussian states instead yield undistillable encoded states that are close to Pauli eigenstates~\cite{pantaleoni2021subsystem}, so a different type of Gaussian state must be used instead.
} 

In a cluster-state setting, there are two straightforward ways to implement this protocol using state injection. The first is to inject externally produced noisy magic states 
using the above or some other method. The second is to inject Gaussian states (notably, vaccum states) and then project them into the GKP code space through teleportation as described in Eq.~\eqref{KnillCartoonQunaught}. 
In either case, homodyne measurements on surrounding cluster-state modes provide the GKP Clifford machinery to perform distillation as needed within the cluster state itself.

An intriguing alternative approach that also yields a distillable magic state is to measure half of an encoded GKP Bell pair using heterodyne detection~\cite{Baragiola2019}, which projects that mode onto the coherent-state basis. 
We modify this approach for streamlined implementation in macronode cluster states. We make use of two facts. First,
beam splitters acting before projections onto coherent states are equivalent to projections onto different coherent states,
\begin{align}
    \bra{\alpha_1} \otimes \bra{\alpha_2} \hat{B}_{12} = 
    \bra{ \alpha_+ } \otimes
    \bra{ \alpha_- } 
    \, ,
\end{align}
with $\alpha_\pm \coloneqq \tfrac{1}{\sqrt{2}} (\alpha_1 \pm \alpha_2)$. With this, beam splitters can be effectively removed by postprocessing of outcomes. 
Second, performing heterodyne on both modes of the single-mode macronode gadget, Eq.~\eqref{circuit:macrnonodegadget}, disentangles the input mode, leaving the final two modes in a GKP Bell pair, Eq.~\eqref{eq:GKPBellPair}:
 \begin{align}\label{circuit:magicmaker}
    \Qcircuit @C=1.75em @R=1.75em 
    {
	&\lstick{\bra{\alpha_a}}   & \bsbal{1} & \qw & \rstick{\text{(in)}} \qw[-1] &  \\
	&\lstick{\bra{\alpha_b}}  & \qw   &  \bsbal{1} & \rstick{\ket{\qunaught}} \qw \\
	&\lstick{\text{(out)}}	& \qw  &   \qw  & \rstick{ \ket{\qunaught} } \qw
	}\, 
	\quad
	\raisebox{-1.7em}{~=} 
	\qquad
    \Qcircuit @C=1.75em @R=1.75em 
    {
	&\lstick{\bra{\alpha_+}}   & \qw & \qw & \rstick{\text{(in)}} \qw[-1] &  \\
	&\lstick{\bra{\alpha_-}}  & \qw   &  \bsbal{1} & \rstick{\ket{\qunaught}} \qw \\
	&\lstick{\text{(out)}}	& \qw  &   \qw  & \rstick{ \ket{\qunaught} } \qw
	}\, 
\quad
	\!\!\raisebox{-2em}{.}	
    \end{align}
Since one of these modes is measured by heterodyne detection, this implements the conditional GKP magic-state approach described above.
From a resource perspective, this approach is equivalent to introducing a vacuum state into the cluster state, since heterodyne can be realized via dual homodyne detection in which vacuum enters through an empty beamp slitter port. 

This technique has broader application, too. When the input state is itself half of a GKP Bell pair, two GKP magic states are produced, which are generally different but both distillable.
In the context of the two-mode macronode gadget, Eq.~\eqref{twoModeCircuit_QRL}, performing heterodyne detection on all measured modes produces up to four GKP magic states---one at each output and one for each input mode 
that is part of a GKP Bell pair with another mode of the cluster state.
The generated GKP magic states are nearby in the cluster state, potentially making their Clifford-circuit distillation convenient.

The suitability of any particular method of including magic states in a cluster-state framework will depend on the details of the architecture. Our contribution here is to illustrate how to implement the methods proposed in Ref.~\cite{Baragiola2019}, which involve either injection of the vacuum or heterodyne detection, in a way that dovetails naturally with QRL-based cluster states.

\section{GKP-qubit gate noise}
\label{sec:squeezingthresholds}

The success of error correction and ultimately the reliability of the computation depend on the amount of noise in the resources used as well as the machinery employed to handle this noise. 
The Clifford resources for ideal GKP quantum computing with macronode-based cluster states generally include 
ideal zero-momentum, zero-position, and GKP qunaught states~\cite{Walshe2020,tzitrin2021fault}. These ideal resources, which are combined using beam splitters and then measured, allow multiple GKP Clifford gates to be applied consecutively without introducing additional noise. Physical approximations to these resource states---i.e.,~momentum- and position-squeezed vacuum states and approximate GKP qunaught states, respectively---have finite energy, so teleported GKP Clifford gates are accompanied by additional noise~\cite{Walshe2020}.
\blk
In this section, we quantify how the CV noise inherent in a macronode cluster state manifests as logical noise on GKP-encoded qubits. Provided the qubit-level noise is low enough, concatenation with qubit codes can reduce the effective noise as low as required
for any particular quantum computation.

Following the method introduced in Ref.~\cite{menicucci2014fault} for noise analysis, we consider teleported CV gates followed by GKP error correction as noisy qubit gates. As we have shown above, the GKP Cliffords in Table~\ref{tab:twoModeMeasurementAngles} followed by GKP error correction can be performed in a single step.
To quantify the performance of these error-corrected gates, we calculate the qubit-level \emph{error rate} associated with each gate after the error correction. This is the probability that one or more of the GKP error correction steps fails, resulting in a qubit-level error.

Having done so, we can abstract away the CV level entirely and treat these gates as noisy qubit gates whose error rates may be compared to the fault-tolerance thresholds of typical quantum error correction codes using qubits to make claims about the level of squeezing required for fault tolerance (before any other sources of error are considered). While it is possible to choose a qubit-level code and numerically derive a squeezing threshold for that specific code~\cite{tzitrin2021fault,larsen2021architecture,Noh2021lowoverhead}, we choose to remain agnostic about the qubit-level code used and instead focus on the gate error rates associated with different levels of squeezing in the approximate states.

As discussed in Sec.~\ref{subsec:magicstates}, supplementing fault-tolerant GKP Clifford operations with easy-to-produce vacuum or thermal states can be used to make GKP magic states required for universality~\cite{Baragiola2019}. The quality requirements for the Clifford gates are much more stringent than for noisy magic states since the latter can be improved using distillation if the Clifford circuits are good enough. For this reason, we focus on the noise in Clifford gate implementations.

\subsection{Gaussian blurring channel}
\label{subsec:gaussblur}

In what follows, we will make liberal use of the (incoherent) Gaussian blurring channel
  \begin{align}\label{GaussianNoiseChannel}
      \mathcal{E}_{\delta^2}
      &\coloneqq \frac{1}{\pi \delta^2}\int d^2 \alpha \, e^{-|\alpha|^2/\delta^2 } \op{D}( \alpha ) \odot \op{D}^\dagger ( \alpha)
      \, ,
  \end{align}
which applies random displacements with amplitudes drawn independently from a zero-mean Gaussian with variance $\delta^2$. 
This channel will be used to conceptually describe the initial states in the analysis---noisy GKP and qunaught states---as well as the noise-accumulation effects of each step in the measurement-based quantum computation. 
The action of the channel on a state is the Gaussian weighted average of displacements; in the Wigner representation, this is simply a Gaussian blurring in each quadrature~\cite{mensen2021}.

Consider a multimode Gaussian state with zero mean and Wigner covariance matrix~$\mat \Sigma$. The elements of this covariance matrix are $\Sigma_{jk} = \tfrac 1 2 \avg{\{\op x_j, \op x_k\}}$, where $\opvec x = (\op q_1, \dotsc, \op q_N, \op p_1, \dotsc, \op p_N)^\tp$, and $\{ \cdot, \cdot \}$ is the anticommutator. This is the same ordering convention for the quadratures used, for instance, in Ref.~\cite{menicucci2014fault}. Applying this channel~$\mathcal{E}_{\delta^2}$ independently on all~$N$ modes produces a new Gaussian state with the same mean and broader Wigner covariance~$\mat \Sigma + \delta^2 \mat \id_{2N}$, where $\mat \id_{2N}$ is the ${2N\times 2N}$ identity matrix.
For a single mode, we write the action of~$\mathcal{E}_{\delta^2}$ as a simple map on the Wigner covariance matrix:
\begin{align}
\label{eq:bluroncovmat}
    \mat \Sigma
+
    \begin{bmatrix}
        \delta^2 & 0 \\
        0 & \delta^2
    \end{bmatrix}
    \xmapsfrom{~\mathcal{E}_{\delta^2}~}
    \mat \Sigma
    \, .
\end{align}
Note that this evolution is right to left.

\subsection{Representing approximate GKP states and teleportation-based error correction}
\label{subsec:repapproxGKPteleportation}

Ideal GKP states are unique in that their Wigner functions are a weighted sum of delta spikes---i.e.,~individual Gaussians with covariances approaching the zero matrix (while remaining positive definite). Formally, we can write this covariance matrix as $0^+ \mat \id_2$, where $0^+$ is an infinitesimal positive constant. The reason such a spike is allowed in a Wigner function---despite violating the Heisenberg uncertainty principle when considered on its own---is that it is part of an infinite ensemble of regularly spaced spikes that form the GKP grid~\cite{GKP,mensen2021}. Finite approximations to these states can have spikes as narrow as allowed by the envelope of the state~\cite{mensen2021}, with larger envelopes allowing for smaller spikes.

Physical approximations to ideal GKP states are described by replacing each $\delta$-function spike in the ideal wave functions (position or momentum) with a sharp Gaussian and then damping the comb with a broad Gaussian envelope~\cite{GKP, matsuura2019, mensen2021}.
For our study of logical error rates and corresponding levels of squeezing, we consider high-quality GKP states. In this regime, the analysis is simplified by ignoring the broad envelope (or broadening it out to infinity) and considering only the noise on the individual spikes~\cite{Glancy2006,menicucci2014fault, noh2020fault, fukui2021all, Hillmann2021performance}.\footnote{This can be modeled formally as resulting approximately from a high-quality physical GKP state (i.e.,~one that has a finite Gaussian envelope that limits the total energy of the state) and twirling it by the GKP stabilizer group~\cite{noh2020fault,mensen2021}. 
This leaves the logical information invariant but blurs out the envelope (in the Wigner picture) to the point where it is approximately constant.} The result is a Gaussian-blurred version of an ideal GKP state, i.e.,~$\mathcal{E}_{\delta^2}(\outprod{\psi_\GKP}{\psi_\GKP})$, which gives a blurred version of the original state,\footnote{Normalization for ideal GKP states is a subtle issue~\cite{GKP,Baragiola2019,noh2020fault,Walshe2020,mensen2021}. Whatever norm is chosen for the ideal state is preserved by this channel.} whose Wigner-function spikes have covariance matrix
    \begin{align} \label{eq:CovarianceSpike}
        \mat{\eta} = \begin{bmatrix} \delta^2 & 0 \\ 0 & \delta^2 \end{bmatrix}
        \, 
    \end{align}
that we refer to as the \emph{error matrix} for the approximate GKP state. This is because it determines the probability of a logical error occurring after ideal GKP error correction is performed on the state~\cite{Glancy2006,menicucci2014fault}.

The diagonal elements of~$\mat \eta$ give the variance along each quadrature that would be measured in an experiment, $(\sigma^2_{\text{spike}, q},\sigma^2_{\text{spike}, p})$. We call these \emph{measured variances} to distinguish them from wave-function variances such as $\Delta^2$ and $\kappa^2$ as defined in Ref.~\cite{GKP} for pure approximate GKP states.%
\footnote{For ease of comparison to other works, we note that these measured variances would also appear in the quadrature statistics of pure approximate GKP states with $\Delta^2 = 2\sigma^2_{\text{spike}, q}$ and $\kappa^2 = 2\sigma^2_{\text{spike}, p}$.
Our input GKP states, which are slightly blurred (and thus, mixed) ideal states, have the same measured quadrature statistics for each spike as pure approximate GKP states with $\Delta^2 = \kappa^2 = 2 \delta^2$.}
The off-diagonal elements of~$\mat \eta$ describe correlations between the two quadratures (\emph{i.e.,}~the covariance of the two). 
For the blurred ideal states under consideration, the variances are
$\sigma^2_{\text{spike}, q} = \sigma^2_{\text{spike}, p} = \delta^2$, and there are no correlations between them (zeros off the diagonal), which gives $\mat \eta$ the form shown.
\blk

In figures and discussion, we present the measured variances of the resources used for quantum computation---the input GKP qubit and qunaught states comprising the cluster sate---in decibels, with the measured vacuum variance of $\frac 1 2$ taken as the reference value. This is a standard way to characterize the quality of the squeezing in both types of states:
    \begin{align}\label{squeezingfactor}
        (\delta^2)_\dB
        = -10\log_{10} ( 2 \delta^2 ).
    \end{align} 
We choose a convention for which $\delta^2 < \tfrac 1 2$ (squeezed below vacuum variance) corresponds to positive decibel values.

Using the noise model described above, we replace the pure states $\ket{\psi}$ and $\ket{\phi}$ in the macronode gadget, Eq.~\eqref{circuit:macrnonodegadget}, with blurred qunaught states, $\mathcal{E}_{\delta^2}(\outprod{\qunaught}{\qunaught})$.
In this more general case, the input state evolution through the circuit is not given by the Kraus-operator map in Eq.~\eqref{eq:Krausmap}; instead, it is described by the map%
\footnote{The sandwiching of $\op{\Pi}_\GKP$ by two instances of $\mathcal{E}_{\delta^2}$ is the result of teleportation through the noisy GKP Bell pair~\cite{fukui2021all}. It is the mixed-state version of the Kraus operator discussed in Ref.~\cite{Walshe2020} that implements GKP error correction when used with pure approximate qunaughts, namely, $e^{-\beta \op n} \op{\Pi}_\GKP e^{-\beta \op n}$. When mixed qunaughts (as discussed here) are used instead, the damping operators $e^{-\beta \op n}$ become quantum channels~$\mathcal{E}_{\delta^2}$. More precisely, twirling $e^{-\beta \op n}$ by the GKP stabilizers gives the channel~$\mathcal{E}_{\delta^2}$~\cite{mensen2021}, with $\beta = 2\delta^2$ under our assumption of high-quality states ($\delta^2 \ll 1$). In a related (but not identical) setting to what we consider here, Ref.~\cite{Hillmann2021performance} compared average fidelity for teleportation error correction using these two types of noisy GKP states and found no differences.}%
    \begin{align}\label{errorChannel}
        \op{\rho}_\text{out} = 
        \op D_{\text{c}}  \mathcal{E}_{\delta^2} \big[ \op{\Pi}_\GKP  \mathcal{E}_{\delta^2}(\op{D}_\mu \op{V} \op{\rho}_\text{in} \op{V}^{\dagger} \op{D}^\dagger_\mu )  \op{\Pi}_\GKP \big]
        \op D_{\text{c}}^\dag
        \, ,
    \end{align}
where $\op{V}$ stands for the intended gate to be applied $\op{V}(\theta_a,\theta_b)$ and $\op{D}_\mu$ stands for the outcome-dependent displacement $\op{D}(\mu_{a,b})$. The final displacement $\op D_{\text{c}}$ implements a possible logical Pauli correction $\bar{P}_{\text{c}}$ at the very end of the protocol, which is determined by a decoder using the measurement outcomes~$\mu_{a,b}$.%
\footnote{
The fact that only logical corrections are needed is a feature of teleportation-based~\cite{Walshe2020} (Knill-style~\cite{Knill2005a}) error correction. The state returns to the GKP subspace (followed by blurring) via the teleportation, but logical errors can be introduced in the process. These are heralded by the measurement outcomes and corrected by applying an appropriate~$\op D_{\text{c}}$. When this correction is incorrectly determined, a logical Pauli error occurs.}%

\subsection{Noise in GKP macronode-based quantum computing}
\label{subsec:noisecalc}

We now focus on using the QRL macronode gadgets for GKP quantum computing, where the input state~$\op \rho_{\text{in}}$ is itself an approximate GKP state from the output of the previous macronode gadget, $\op \rho_{\text{out,previous}}$. 
Since  displacements commute with~$\mathcal{E}_{\delta^2}$, we can postpone $\mathcal E_{\delta^2}$ to the very end and collect $\op D_\text{c} \op \Pi_{\GKP} \op D(\mu_{a,b})$ into a single operation that describes \emph{ideal} GKP error correction.

With this change of ordering, we schematically describe the map implemented in Eq.~\eqref{errorChannel} as a sequence of transformations on the input state $\op{\rho}_\text{in}$,
\begin{align}\label{eq:rhoevolution}
    \op \rho_{\text{out}}
    \xmapsfrom{~\mathcal{E}_{\delta^2}~}
    \op \rho_3
    \xmapsfrom{~\text{GKP EC}~}
    \op \rho_2
    \xmapsfrom{~\mathcal{E}_{\delta^2}~}
    \op \rho_1
    \xmapsfrom{~\op{V}(\theta_a,\theta_b)~}
    \op \rho_{\text{in}}
    \, ,
\end{align}
where ``GKP~EC'' stands for $\op D_{\text{c}} \op \Pi_{\GKP} \op D(\mu_{a,b})$ and includes the final logical Pauli correction, if required. 
This sequence is read right to left, with $\op \rho_{1,2,3}$ mathematically representing the state at various points through the evolution.
Although we have broken it down into four pieces for analysis, the entire transformation $\op \rho_{\text{in}} \mapsto \op \rho_{\text{out}}$ actually happens all at once
during teleportation through a single macronode (which includes the final displacement correction). That is, this entire operation is performed in a \emph{single step} of teleportation. We stress this fact to contrast with prior protocols in which the gates and GKP error correction are performed separately~\cite{Larsen2020noiseAnalysis,larsen2021architecture}.
\blk

Since the input state~$\op \rho_{\text{in}}$ is itself an approximate GKP state,
we can study the noise properties of macronode gadget evolution
by evolving the input-state error matrix under an analogous transformation,
\begin{align}\label{eq:etaevolutionsimple}
    \mat \eta_{\text{out}}
    \xmapsfrom{~\mathcal{E}_{\delta^2}~}
    \mat \eta_3
    \xmapsfrom{~\text{GKP~EC}~}
    \mat \eta_2
    \xmapsfrom{~\mathcal{E}_{\delta^2}~}
    \mat \eta_1
    \xmapsfrom{~\op{V}(\theta_a, \theta_b)~}
    \mat \eta_{\text{in}}
    \, ,
\end{align}
with the understanding that $\mat \eta_s$ is the corresponding error matrix for $\op \rho_s$ for each subscript~$s$ in Eq.~\eqref{eq:rhoevolution}.
Also, note that displacements have no effect on error matrices, which allows us to reuse the same notation for them as above. In what follows, we illustrate the effect of each of these operations on the error matrix in order to determine the success probability for a variety of GKP logical Clifford gates, which will ultimately depend on the amount of initial squeezing~$(\delta^2)_\dB$.

The Gaussian unitary operation $\op{V}(\theta_a, \theta_b)$, which implements a GKP Clifford gate using the measurement angles in Table~\ref{tab:twoModeMeasurementAngles}, updates the error matrix according to a symplectic matrix $\mat{S}_{\op{V}}$ representing the Heisenberg action of the gate on the quadratures~\cite{menicucci2014fault}. Thus,
\begin{align}
    \mat \eta_1
&=
    \mat{S}_{\op{V}}
    \mat \eta_{\text{in}}
    \mat{S}_{\op{V}}^\tp
    \, .
\end{align}
The effect of~$\mathcal{E}_{\delta^2}$ is additive on the error matrix, Eq.~\eqref{eq:bluroncovmat}, so
\begin{align}
    \mat \eta_2
&=
    \mat \eta_1
+
    \begin{bmatrix}
        \delta^2 & 0 \\
        0 & \delta^2
    \end{bmatrix}
    \, .
\end{align}
Ideal GKP error correction produces an output state with delta-function spikes. Thus, formally, $\mat \eta_3 \to 0^+ \mat \id$. This is a good time to recall that this error matrix is never realized in practice and is merely a mathematical tool used to assist with the calculation. Finally, the second $\mathcal{E}_{\delta^2}$ gives a fixed, final (and physical) error matrix of
\begin{align}
    \mat \eta_{\text{out}}
&=
    \begin{bmatrix}
        \delta^2 & 0 \\
        0 & \delta^2
    \end{bmatrix}
    \, .
\end{align}
After the whole procedure, the noise properties of the output state are identical to those of the qunaught states used for error correction. However, during the error correction, logical errors may have been introduced, and it is these errors that constitute the logical-qubit gate noise.
\blk

Noting that it is $\op \rho_2$ that undergoes GKP error correction, $\mat \eta_2$ can be used to determine its probability of success. Furthermore, this error matrix depends on $\op V(\theta_a, \theta_b)$, so for these reasons, we rename $\mat \eta_2$ as
\begin{align}
    \mat \eta_{\op V}
\coloneqq
    \mat \eta_2
=
    \mat{S}_{\op V}
    \mat \eta_{\text{in}}
    \mat{S}_{\op V}^\tp
+
    \begin{bmatrix}
        \delta^2 & 0 \\
        0 & \delta^2
    \end{bmatrix}
    \, .
\end{align}
This notation will let us differentiate between the error matrices for different gates.

We also need the analogous result for a two-mode gate, corresponding to the output of Eq.~\eqref{eq:twomodekraus_QRL}. We write $\mat \eta^{(2)}$ for a general two-mode error matrix, which is the covariance matrix for a single spike in the Wigner function of a two-mode GKP state, with respect to the quadrature ordering~$(\op q_1, \op q_2, \op p_1, \op p_2)$. An entirely analogous procedure to the single-mode case gives the relevant error matrix
\begin{align}
    \mat \eta_{\op V}^{(2)}
\coloneqq
    \mat \eta_2^{(2)}
=
    \mat{S}_{\op V}
    \mat \eta_{\text{in}}^{(2)}
    \mat{S}_{\op V}^\tp
+
    \begin{bmatrix}
        \delta^2 & 0 & 0 & 0\\
        0 & \delta^2 & 0 & 0\\
        0 & 0 & \delta^2 & 0\\
        0 & 0 & 0 & \delta^2
    \end{bmatrix}
    \, .
\end{align}

The error matrices for the single-mode gates in Table~\ref{tab:twoModeMeasurementAngles} are
\begin{align}\label{singleModeErrorMats}
\mat{\eta}_{\op{I}} = \mat{\eta}_{\op{F}}
    &=\begin{bmatrix}  
        2 \delta^2  & 0  \\
        0           & 2 \delta^2   
    \end{bmatrix} 
    \, , \quad \quad 
    \mat{\eta}_{\op{P}(\pm 1)}
    =\begin{bmatrix}  
        2 \delta^2  & \delta^2  \\
        \delta^2           & 3 \delta^2   
    \end{bmatrix}
 \, .
 \end{align}
For the two-mode gate $\CZ(\pm 1)$ that completes the Clifford generating set, the error matrix is 
\begin{align}\label{CzErrorMats}
\mat{\eta}_{\CZ(\pm 1)}
&=\begin{bmatrix}  2 \delta^2  & 0 & 0 & \delta^2  \\
 0 & 2 \delta^2  & \delta^2  & 0 \\
 0 & \delta^2  & 3 \delta^2  & 0 \\
 \delta^2  & 0 & 0 & 3 \delta^2  \\ 
 \end{bmatrix}
 \, .
 \end{align}
The error matrix for the SWAP gate is
\begin{align}\label{SwapErrorMats}
\mat{\eta}_\text{SWAP}
&=\begin{bmatrix}  2 \delta^2  & 0 & 0 & 0  \\
 0 & 2 \delta^2  &0  & 0 \\
 0 &0  & 2 \delta^2  & 0 \\
0  & 0 & 0 & 2 \delta^2  \\ 
 \end{bmatrix}
 \, .
 \end{align}
For comparison, we also consider the separable gates. The error matrices for two-mode identity and identical Fourier transforms on both modes are the same as that for the SWAP gate in Eq.~\eqref{SwapErrorMats}:
 \begin{align}
  \mat{\eta}_{\op I \otimes \op I}&=
  \mat{\eta}_{\op F \otimes \op F}=
  \mat{\eta}_\text{SWAP}
    \, .
 \end{align}
Finally, the error matrices for identical unit-weight shears on both modes are
 \begin{align}    
  \mat{\eta}_{ \op P(\pm 1) \otimes \op P(\pm 1)}&=\begin{bmatrix}  2 \delta^2  & 0 & \delta^2 & 0  \\
 0 & 2 \delta^2  & 0  & \delta^2 \\
 \delta^2 & 0  & 3 \delta^2  & 0 \\
 0  & \delta^2 & 0 & 3 \delta^2  \\ \end{bmatrix}
 \, .
\end{align}
We will use these error matrices to determine gate error rates as a function of $(\delta^2)_\dB$ in the next subsection.
\blk

\subsection{Logical gate error rates and fault tolerance}
\label{results}
The probability that a logical Pauli error is introduced during error correction is a function of the noise in the input GKP data qubit, the noise in the qunaught states that comprise the macronode gadget, and which Clifford is being implemented. The error matrices above capture all of these effects on each Wigner-function spike of a GKP state. For the square-lattice GKP code, the probability that a Pauli-$X$ or a Pauli-$Z$ error is introduced during error correction is determined by the leakage of the Wigner-function GKP spike out of its unit cell in $q$ and in $p$, for each half of error correction~\cite{menicucci2014fault}. These logical error probabilities are given by 
    \begin{align}\label{probLogicalErr}
        P_{\text{err,$X|Z$}}=1 - P_{\text{succ,$q|p$}},
    \end{align}
where $\cdot | \cdot$ represents alternatives, respectively, on each side of the equation.
The probability of success (no error during that half of error correction) is given by 
    \begin{align}\label{probSuccMode}
        P_{\text{succ,$q|p$}} = \erf\left(\sqrt{\frac{\pi}{8 \, \sigma^2_{\text{spike},q|p}}}\right),
    \end{align}
where $\sigma^2_{\text{spike},q|p}$ is the measured variance of the GKP spike, 
either $q$ or $p$~\cite{menicucci2014fault}.
For each gate, we find these spike variances from the diagonal elements of the gate's noise matrix (which are all integer multiples of the baseline noise $\delta^2$ in the input GKP and qunaught states).

To consider gate errors at the logical-qubit level, we are interested in the probability that at least one Pauli error occurs,
    \begin{align}\label{probErr}
        P_{\text{err}}= 1 - P_{\text{succ}},
    \end{align}
where $P_\text{succ}$ is the total success probability, given by
    \begin{align}
        P_\text{succ} = \prod_j P^{(j)}_{\text{succ},q} P^{(j)}_{\text{succ},p}
        \, ,
    \end{align}
with $j$ iterating over the modes being considered (either one or two modes, depending on whether we are considering single- or two-mode gates).

\begin{figure}[!t]
    \centering
    \includegraphics[width=\columnwidth]{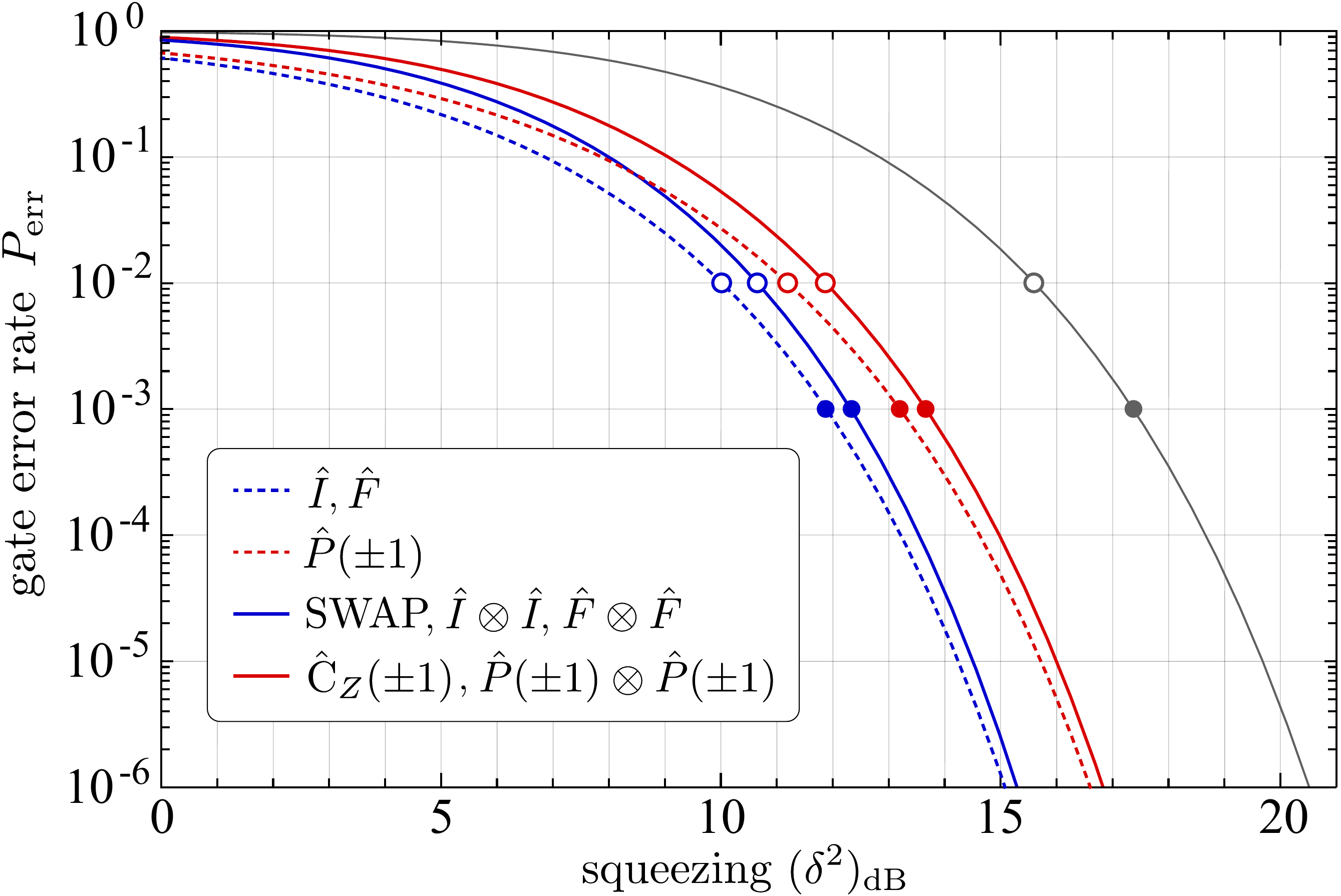}
    \caption{
    Logical gate error rates, Eq.~\eqref{probErr}, for resources (input GKP states and qunaught states) of a given quality described by the squeezing, Eq.~\eqref{squeezingfactor}. Dashed curves are single-mode gates, and solid curves are two-mode gates. Unfilled (filled) circles indicate the squeezing required for error rate $10^{-2}$ ($10^{-3}$); squeezing values at these points are given in Table~\ref{thresholdComparison}. 
The two-mode gate $\CZ(\pm 1)$ sets the relevant gate error rate for Clifford implementation, since it is the worst performing gate in the set, Eq.~\eqref{eq:gateconnections}. 
The gray line (furthest right) shows the error rate for the $\CZ$ gate implemented in a canonical CV cluster state using the method of Ref.~\cite{menicucci2014fault}. That method required at least four measurements per mode to do the gate and two additional ones per mode to implement error correction (six measurements per mode in total). Our results, in contrast, require only two measurements per mode in total, which implement both the gate and the error correction simultaneously. This reduction in measurements is the source of the substantial improvement in error rates in our results over those in Ref.~\cite{menicucci2014fault}. A similar reduction in measurements per mode also explains the (more modest) improvement over those in Refs.~\cite{larsen2021architecture,Larsen2020noiseAnalysis}, shown in Table~\ref{thresholdComparison}.
    }
    \label{Fig-sec:squeezingthresholds}
\end{figure}

In Fig.~\ref{Fig-sec:squeezingthresholds}, we plot the gate error rate, Eq.~\eqref{probErr}, as a function of the squeezing in the resources---specifically, the input GKP states and the qunaught states that comprise the macronode cluster state---required to implement the gate. 
Noisy GKP Clifford gates have different error rates, with the worst-performing gate setting the required squeezing for a fixed tolerable error rate. 
That gate is the two-mode gate $\CZ (\pm 1)$ for all levels of squeezing. 

Required squeezing for selected error rates is shown in Table~\ref{thresholdComparison}.
We find that error rates of $10^{-2}$--$10^{-3}$, which are compatible with the thresholds of 3D topological codes ($\sim$1\% for local noise~\cite{Raussendorf2007,Fowler2012surface}), require a minimum of 11.9--13.7~dB of squeezing in the resources. 
For comparison, we also give the required squeezing at these error rates for several previous proposals.

An important benchmark is the required squeezing set out in Ref.~\cite{menicucci2014fault} for canonical CV cluster states and GKP ancilla resources. That work found, using an extremely conservative logical error rate of $10^{-6}$, that with 20.5~dB of squeezing in the resource states (cluster state and GKP states) Clifford gates had no more than this level of error. This level of squeezing 
has been considered a ``squeezing threshold'' associated with the $10^{-6}$ error rate.
We do not use the term ``squeezing threshold'' to characterize the results in this paper to avoid making claims about the practical viability of fault-tolerant quantum computing with a given level of squeezing; doing that would require simulations of particular implementations, such as in Refs.~\cite{larsen2021architecture,bourassa2021blueprint,tzitrin2021fault}. 
Instead, we focus on the relationship between the level of required squeezing to achieve target gate error rates.

\begin{table}[t]
\resizebox{\columnwidth}{!}
{
\def\arraystretch{1.5}
\begin{tabular}{|c|c|c|c|c|c|c|c|c|}
\hline
Gate    & \multicolumn{4}{c|}{Error rate: $10^{-2}$} & \multicolumn{4}{c|}{Error rate: $10^{-3}$} \\ \hline
 & Ref.~\cite{menicucci2014fault}      &Ref.~%
 \cite{Larsen2020noiseAnalysis} & Ref.~\cite{larsen2021architecture} & ~ours~
 & Ref.~\cite{menicucci2014fault} & Ref.~\cite{Larsen2020noiseAnalysis}    & Ref.~\cite{larsen2021architecture} & ~ours~ \\ \hline
$\op I$    &    14.0   &  13.2   &   11.8   &   10.0     &     15.9     &   15.0    &   13.6    &  11.9   \\ \hline
$\op F$    &      14.8 &  14.9   &    11.8 &  10.0   &    16.8 &      16.7         &    13.6    & 11.9    \\ \hline
$\op P(\pm1)$ &     14.4 & 15.2   &  12.5  &   11.2   &       16.4 &   17.1     &   14.5   &  13.7   \\ \hline
$\CZ(\pm 1)$   &    15.6 &  -   & -      &  11.9  &       17.4 &     -     &    -     & 13.7  \\ \hline
$\op F \op F \CZ$   & -  & 16.0     & 13.2    & -     &       -    &  17.6     &    15.0     & -    \\ \hline
\end{tabular}
}
\caption{Squeezing requirements (reported in~dB) for implementing GKP Clifford gates with $10^{-2}$ and $10^{-3}$ logical-qubit error rates. 
We consider the following CV gates that together generate the GKP Clifford group: $\{\op I,\op F,\op P(\pm 1),\CZ(\pm 1)\}$.
We compare the results in this paper (`ours') with those of prior studies~\cite{menicucci2014fault, Larsen2020noiseAnalysis, larsen2021architecture}, noting that we use $\op F \op F \CZ$ as shorthand for $(\op F \otimes \op F) \CZ(1)$ from Ref.~\cite{Larsen2020noiseAnalysis} and $(\op F^\dagger \otimes \op F) \CZ(1)$ from Ref.~\cite{larsen2021architecture}.
}\label{thresholdComparison}
\end{table}

Reference~\cite{menicucci2014fault} employed canonical CV cluster states for ease of theoretical analysis, but these are not the type of cluster state that admit scalable generation. All large-scale cluster states demonstrated to date are based on macronodes~\cite{Yokoyama2013,Asavanant2019,Larsen2019,Yoshikawa2016}, which is why a macronode-based analysis is important. The first reported required squeezing values for macronode cluster states (built with beam splitters) were given in Larsen \emph{et al.}~\cite{Larsen2020noiseAnalysis}, which considered quantum computation protocols for a variety of macronode lattices, including the QRL. 
These protocols differ from ours in two significant ways: their single-mode Clifford gate implementations require two steps---four homodyne measurements---and uses ancilla-assisted GKP error correction (Steane-type error correction~\cite{Steane1997}).
The required squeezing was later improved by 2.8~dB in Larsen \emph{et~al.}~\cite{larsen2021architecture} by upgrading to teleportation-based error correction (Knill-type~\cite{Knill2005a}), yet the new protocol is still limited by the fact that it requires two steps, one to perform the gate and another for GKP error correction.\footnote{That work did not report required squeezing values. We have calculated them using the methods described here, including for the two-mode gate that arises from a different macronode teleportation gadget.}
Our further improvement of $\sim$1.3~dB over that scheme results from combining Clifford gates and teleportation-based GKP error correction into a single teleportation step
. 
Moreover, we have reduced GKP gate implementations to the minimum number of noisy ancilla states in the macronode gadgets (four qunaughts in the two-mode gadget), so further improvements 
are unlikely without devising new schemes for gates requiring fewer ancillae.

\subsection{Applications for fault tolerance in topological codes}
\label{subsec:topological}

A promising avenue for fault-tolerant quantum computing is concatenation of GKP qubits with a topological code~\cite{vuillot2019quantum,hanggli2020enhanced,noh2020fault}. 
Measurement-based implementations wire up cluster states in various ways into 3D cluster states with topological fault tolerance. A key example is the RHG 3D lattice~\cite{Raussendorf2007}, which can be used with GKP qubits~\cite{fukui2020temporal,bourassa2021blueprint} including macronode-based architectures~\cite{larsen2021architecture, tzitrin2021fault}. Since the RHG lattice is of uniform degree 4, it is directly compatible with a QRL construction (i.e.,~with four modes per macronode), as one recent proposal illustrates~\cite{tzitrin2021fault}.

The QRL construction that we consider here serves as a canonical base case that showcases improved use of noisy GKP resources in a macronode setting. Preliminary results indicate that GKP computing with other similar macronode lattices of interest---including 
that employed in Refs.~\cite{larsen2021architecture,Noh2021lowoverhead}---performs equally well. That is, GKP Clifford gates and error correction can be executed in a single step (using the minimum number of noisy qunaught ancilla states) and the logical error rates are identical to those for the QRL presented here. This noise equivalence 
dovetails with current proposals for topological quantum computing by enabling known fault-tolerance thresholds of 10.2~dB in Ref.~\cite{larsen2021architecture} and 9.9~dB in Ref.~\cite{Noh2021lowoverhead}. In those works, simultaneous GKP gate and error correction were anticipated, but the methods to execute them were unknown. Further threshold improvements may result from combining other techniques such as ``hyper-enriching'' the GKP qubits in the cluster state~\cite{tzitrin2021fault}, concatenating with other codes~\cite{fukui2021efficient}, and using the analog syndrome to improve the error correction~\cite{fukui2018high,yamasaki2020polylog}.

\section{Conclusion}
\label{con}

Using the quad-rail lattice CV cluster state,
we have introduced improvements that allow for more efficient use of the cluster state for fault-tolerant quantum computing with the GKP code. 
These improvements include single-step implementations of all single- and two-mode GKP Clifford gates using only the minimum number of noisy ancilla states in the QRL macronode gadgets that realize the gates.
This allows GKP Cliffords and GKP error correction to be performed simultaneously, lowering GKP gate noise by at least 1.3~dB over similar protocols.
Additionally, we show how to produce GKP magic states without modifying the cluster state itself by using heterodyne detection instead of homodyne to measure the modes.

It is interesting to note that for a squeezing of 10.5~dB (the fault-tolerance threshold reported in Ref.~\cite{bourassa2021blueprint}), we find a gate error rate of 3.6\%---very near that of the RHG lattice under local noise (3.3\%~\cite{Raussendorf2007}), which is the code that paper uses for concatenation.
However, beyond simple comparisons that can give insight for rigorous studies, one should resist the temptation to take the gate-error rates reported here and draw conclusions about fault tolerance. Fault-tolerance thresholds depend on many specific factors, notably the decoder and the error model. 

Finally, preliminary studies indicate that the results here are not unique to the QRL construction: simultaneous GKP Clifford gates and error correction can be implemented in various other macronode lattices with identical gate noise to that for the QRL presented here. This puts many macronode cluster states on the same footing, providing flexibility for experimental implementation. This will be reported in a future publication.

\acknowledgments

We thank Mikkel Larsen for useful feedback. This work is supported by the Australian Research Council Centre of Excellence for Quantum Computation and Communication Technology (Project No.\ CE170100012).
BQB is additionally supported by the Japan Science and Technology Agency through the Ministry of Education, Culture, Sports, Science, and Technology Quantum Leap Flagship Program.

\appendix
\numberwithin{equation}{section}
\renewcommand{\theequation}{\thesection\arabic{equation}}

\section{Double bouncing 
beam splitters}\label{appendix:symplecticmatrices}

Bouncing is the mathematical relationship that allows us to move a quantum operation from one mode to another when those two modes are prepared in a maximally entangled state $\ket{\EPR}$. For some single mode operator $\op{O}_1$ on mode 1, the bounce relation is
    \begin{align}
        \op{O}_1 \otimes \op{I}_2\ket{\EPR}_{12} = \op{I}_1 \otimes \op{O}^\tp_2 \ket{\EPR}_{12}
        \,
    \end{align}
where the transpose~$^\tp$ is taken in the position basis for the EPR state defined in Ref.~\cite{Walshe2020}. 
More details and the explicit bouncing of various single-mode operators are described in Ref.~\cite{Walshe2020}. Here, we ``double bounce'' several useful Gaussian two-mode gates. 
We begin with a two-mode gate between a pair of modes that are each one half of separate EPR states. Double bouncing refers to bouncing the two-mode gate through both EPR states (single bouncing would be through just one of the EPR states, but we do not consider that case here). Examples follow that make this concept clear.

\subsection{Controlled-shift gates}
The first useful relation is the double bounce of a CV controlled-$Z$ gate of weight $g$, Eq.~\eqref{eq:CVCZ}:
  \begin{align}
        \CZ^{24}(g) \ket{\EPR}_{12} \otimes \ket{\EPR}_{34} = \CZ^{13}(g) \ket{\EPR}_{12} \otimes \ket{\EPR}_{34}
        \, ,
    \end{align}
where we use superscripts to denote the modes on which the gate acts.
This relation is found easily, because the generator for $\CZ(g)$ is a tensor-product operator, $\op{q} \otimes \op{q}$, and the transpose (in the position basis) of each single-mode position operator is trivial; $\op{q}^\tp \rightarrow \op{q}$. The circuit diagram for this relation is
	\begin{align}\label{}
	\begin{split}
        \Qcircuit @C=1.1em @R=1.25em {
            &\qw &\qw      &\qw[-1] \ar @{-} [dr(0.5)] &   \\
	        &\qw &\ctrlg{\raisebox{1.25em}{\scriptsize{$g$}}}{2} &\qw[-1] \ar @{-} [ur(0.5)] &   \\
	        &\qw &\qw      &\qw[-1] \ar @{-} [dr(0.5)] &   \\
	        &\qw &\ctrl{0}    &\qw[-1] \ar @{-} [ur(0.5)] &  
		}
		\quad \raisebox{-2em}{$=$} \quad
		\Qcircuit @C=1.1em @R=1.25em {
            &\qw &\ctrlg{\raisebox{1.25em}{\scriptsize{$g$}}}{2} &\qw[-1] \ar @{-} [dr(0.5)] &   \\
	        &\qw &\qw      &\qw[-1] \ar @{-} [ur(0.5)] &   \\
	        &\qw &\ctrl{0} &\qw[-1] \ar @{-} [dr(0.5)] &   \\
	        &\qw &\qw      &\qw[-1] \ar @{-} [ur(0.5)] &  
		}
	\end{split}	
	\, .
    \end{align}
This technique can be used to double bounce more general controlled shift gates along any quadrature, because of the fact that $[\op{R}(\theta)]^\tp = \op{R}(\theta)$, and this gate can be used to change the quadrature basis for the target mode of the controlled shift. The next example illustrates this.

Here we double bounce a CV controlled-$X$ gate, defined for control mode~$j$, target mode~$k$, and real weight~$g$, as
    \begin{align} \label{eq:CVCX}
        \CX^{jk}(g) &\coloneqq e^{- i g \op q_j \otimes \op p_k} = \op{F}^\dagger_k \CZ[g] \op{F}_k
        \, .
    \end{align}
From the result for $\CZ(g)$ above, double bouncing a CV controlled-$X$ gate is straightforward because the two are related by Fourier transforms that bounce trivially since $\op{F}^\tp = \op{F}$~\cite{Walshe2020}. Decomposing a controlled-$X$ gate and bouncing the operators one by one gives
  \begin{align}
        \CX^{24}(g) \ket{\EPR}_{12} \otimes \ket{\EPR}_{34} = \CX^{31}(-g) \ket{\EPR}_{12} \otimes \ket{\EPR}_{34}
        \,
    \end{align}
with a notable change of sign in the weight, $g \rightarrow -g$, that arises from the Fourier transforms acting in the opposite order after the bounce. The circuit description is
	\begin{align}\label{}
	\begin{split}
        \Qcircuit @C=1.1em @R=1.5em {
            &\qw &\qw      &\qw[-1] \ar @{-} [dr(0.5)] &   \\
	        &\qw &\ctrlg{\raisebox{1.8em}{\scriptsize{$g$}}}{2} &\qw[-1] \ar @{-} [ur(0.5)] &   \\
	        &\qw &\qw      &\qw[-1] \ar @{-} [dr(0.5)] &   \\
	        &\qw &\targ    &\qw[-1] \ar @{-} [ur(0.5)] &  
		}
		\quad \raisebox{-2.5em}{=} \quad
		\Qcircuit @C=1.1em @R=1.5em {
           &\qw & \ctrlg{\raisebox{1.8em}{\scriptsize{$-g$}}}{2}  &\qw[-1] \ar @{-} [dr(0.5)] &   \\
	        &\qw &\qw       &\qw[-1] \ar @{-} [ur(0.5)] &   \\
	        &\qw &\targ     &\qw[-1] \ar @{-} [dr(0.5)] &   \\
	        &\qw &\qw       &\qw[-1] \ar @{-} [ur(0.5)] &  
		}
	\end{split}	
	\, .
    \end{align}

\subsection{Beam splitter}
Double bouncing a beam splitter cannot be performed using the simple rules in Ref.~\cite{Walshe2020} because beam splitter unitaries are not generated by a single tensor product of quadrature operators. However, we can decompose a beam splitter into a product of such operators (up to an overall phase) using a lower-diagonal-upper (LDU) decomposition~\cite{Walshe2020}.
We consider a general beam splitter between modes $j$ and $k$ parametrized by $\theta$,
    \begin{equation} \label{beam splitterdefgen}
        \bsop_{jk}(\theta) \coloneqq e^{-i \theta (\op{q}_j \otimes \op{p}_k - \op{p}_j \otimes \op{q}_k )}
        \, .
    \end{equation}
which has unequal transmissivity $T$ and reflectivity $R$: $T = \cos^2 \theta$ and $R = \sin^2 \theta$, with $T^2 + R^2 = 1$. The 50:50 beam splitter used in the main text, Eq.~\eqref{beam splitterdef}, has $\theta = \pi/4$. 
Hermitian conjugation is equivalent to exchanging the inputs: $\op B_{jk}^\dag(\theta) = \op B_{kj}(\theta)$. 

The LDU decomposition of Eq.~\eqref{beam splitterdefgen} is (up to a global phase),
 \begin{equation} \label{beam splitterLDU}
        \bsop_{jk}(\theta) = \CX^{jk}(\tan \theta) \big[ \op{S}_j^\dagger(\sec \theta) \otimes \op{S}_k(\sec \theta) \big] \CX^{kj}(-\tan \theta)
        \, ,
    \end{equation}
which is represented by the circuit diagram
\begin{equation}\label{BsLdu}
\begin{split}
         \Qcircuit @C=1em @R=3.6em @! 
         {
         	& \varbs{1} & \qw \\
         	& \qw       & \qw
  		  }
\quad \raisebox{-1.7em}{$=$} \quad
    \Qcircuit @C=1em @R=2em {
         	& \ctrlg{\raisebox{0em}{\scriptsize{$\tan \theta$}}}{1}  & \gate{S^\dag(\sec \theta)} & \targ                                                   & \qw \\
         	& \targ                                                  & \gate{S(\sec \theta)}      & \ctrlg{\raisebox{0em}{\scriptsize{$-\tan \theta$}}}{-1} & \qw
  		  }
    \end{split} 
    \, .
	\end{equation}
\blk

We double bounce each of these gates one by one, noting that $[\op{S}(r)]^\tp = \op{S}^\dagger(r)$. 
The result is the relation,
    \begin{equation}
        \bsop_{24}(\theta) \ket{\EPR}_{12} \otimes \ket{\EPR}_{34} = \bsop_{31}(\theta) \ket{\EPR}_{12} \otimes \ket{\EPR}_{34}
    \end{equation}
described by the circuit diagram
	\begin{align}\label{eq:doublebounceBS}
	\begin{split}
    \Qcircuit @C=1.2em @R=1.5em {
        &\qw &\qw        &\qw[-1] \ar @{-} [dr(0.5)] &   \\
	    &\qw &\varbs{2}[\raisebox{1.3em}{\scriptsize{$\theta$}}]  &\qw[-1] \ar @{-} [ur(0.5)] &   \\
	    &\qw &\qw        &\qw[-1] \ar @{-} [dr(0.5)] &   \\
	    &\qw &\qw        &\qw[-1] \ar @{-} [ur(0.5)] &  
		}
		\quad \raisebox{-2.1em}{$=$} \quad
		\Qcircuit @C=1.2em @R=1.5em {
        &\qw &\qw        &\qw[-1] \ar @{-} [dr(0.5)] &   \\
	    &\qw &\qw        &\qw[-1] \ar @{-} [ur(0.5)] &   \\
	    &\qw &\varbs{-2}[\raisebox{-1.4em}{\scriptsize{$\theta$}}] &\qw[-1] \ar @{-} [dr(0.5)] &   \\
	    &\qw &\qw        &\qw[-1] \ar @{-} [ur(0.5)] &  
		}
		\end{split}
		\, .
    \end{align}

\section{Derivation of the two-mode gate for the QRL} \label{TwoModeV}
\red \blk

Any two beam splitters that share only one wire do not commute. We can, however, use the \emph{four-splitter identity} to commute pairs of beam splitters. The identity, introduced in Ref.~\cite{alexander2016flexible}, is
\begin{equation}\label{FourSplitter}
\begin{split}
\centering
\Qcircuit @C=1.5em @R=1.75em {
    &\qw&\qw&\bsbal{1}&\qw\\
    &\qw&\bsbal{1}[-->]&\qw&\qw\\
    &\qw&\qw&\bsbal{1}&\qw\\
    &\bsbal{-3}[-->]&\qw&\qw&\qw
}
\quad \raisebox{-2.5em}{=} \quad
\Qcircuit @C=1.5em @R=1.75em {
    &\bsbal{1}&\qw&\qw&\qw\\
    &\qw&\qw&\bsbal{1}[-->]&\qw\\
    &\bsbal{1}&\qw&\qw&\qw\\
    &\qw&\bsbal{-3}[-->]&\qw&\qw
}
\end{split} \quad .
\end{equation}
Note that while the two dashed beam splitters commute trivially with each other as do the two black ones, any single pair of a black and a dashed beam splitter alone does not. Nevertheless, the two dashed ones commute with the two black ones as shown.

Here we derive the two-mode gate $\op{V}^{(2)}_\QRL$, Eq.~\eqref{VtwoMode_QRLcircuit}, for the QRL~\cite{alexander2016flexible}. The circuit for the QRL two-mode macronode gadget is shown in Eq.~\eqref{twoModeCircuit_QRL}. First, we use the foursplitter identity, Eq.~\eqref{FourSplitter}, to rearrange the four leftmost beam splitters into the following circuit,
\begin{equation}\label{eq:macronodecircuit4split}
    \begin{split}
\Qcircuit @C=1.4em @R=1.75em {
		&&& \lstick{\brasub{m_a}{p_{\theta_a}}}& \bsbal{1}  &  \qw & \bsbal{3}   & \qw &\qw& \rstick{\text{(in)}} \\
		&&& \lstick{\brasub{m_b}{p_{\theta_b}}} & \qw & \bsbal{3}  &\qw   & \bsbal{1} & \qw & \rstick{\ket{\psi}}& \\
		 \lstick{\text{(out)}}& \qw& \qw & \qw& \qw &\qw & \qw & \qw & \qw& \rstick{\ket{\phi}}& \\
		&&& \lstick{\brasub{m_c}{p_{\theta_c}}}& \bsbal{1} & \qw  & \qw  & \qw & \qw&\rstick{\text{(in)}} \\
		&&& \lstick{\brasub{m_d}{p_{\theta_d}}}  &\qw & \qw & \qw & \bsbal{1}&\qw & \rstick{\ket{\psi'}}& \\
		 \lstick{\text{(out)}} & \qw& \qw&\qw  & \qw &\qw & \qw & \qw&\qw & \rstick{\ket{\phi'}}& \
}
\end{split}
\quad .
\end{equation}
Then, we use the same circuit identities as in Eq.~\eqref{circuit:macrnonodegadget}, which give single-mode $\op{V}$ gates on modes $a$ and $c$, and we pull out the teleported gates (details in Ref.~\cite{Walshe2020}). Together, these give the equivalent circuit,
\begin{equation}\label{}
    \begin{split}
\Qcircuit @C=1.1em @R=1.1em {
		&&\EPRdl& \gate{D(\mu_{a,b})}& \gate{V(\theta_a,\theta_b)}  &  \qw & \bsbal{3}   & \qw & \rstick{\text{(in)}} \\
		&&\EPRul& \qw & \qw & \bsbal{3}  &\qw   & \qw  & \EPRdr& \\
		& \lstick{\text{(out)}}& \qw & \qw& \gate{A(\psi,\phi)} &\qw & \qw & \qw & \EPRur& \\
		&&\EPRdl& \gate{D(\mu_{c,d})}& \gate{V(\theta_c,\theta_d)} & \qw  & \qw  & \qw &\rstick{\text{(in)}} \\
		&&\EPRul& \qw  &\qw & \qw & \qw & \qw & \EPRdr& \\
		& \lstick{\text{(out)}} & \qw&\qw  & \gate{A(\psi',\phi')} &\qw & \qw & \qw & \EPRur& \
}
\end{split}
\quad .
\end{equation}
The next step requires us to move the beam splitter between modes $b$ and $d$ onto modes $a$ and $c$ (mode labels are specified in the measurement bras of circuit~\eqref{eq:macronodecircuit4split}). We double-bounce the left beam splitter using Eq.~\eqref{eq:doublebounceBS} to get
\begin{equation}\label{}
    \begin{split}
\Qcircuit @C=1.1em @R=1.1em {
		&&\EPRdl&\qw& \gate{D(\mu_{a,b})}& \gate{V(\theta_a,\theta_b)}  &  \bsbal{3} & \qw   & \qw & \rstick{\text{(in)}} \\
		&&\EPRul&\qw& \qw & \qw & \qw  &\qw   & \qw  & \EPRdr& \\
		& \lstick{\text{(out)}}&\qw& \qw & \qw& \gate{A(\psi,\phi)} &\qw & \qw & \qw & \EPRur& \\
		&&\EPRdl&\bsbal{-3}& \gate{D(\mu_{c,d})}& \gate{V(\theta_c,\theta_d)} & \qw  & \qw  & \qw &\rstick{\text{(in)}} \\
		&&\EPRul& \qw&\qw  &\qw & \qw & \qw & \qw & \EPRdr& \\
		& \lstick{\text{(out)}} & \qw&\qw&\qw  & \gate{A(\psi',\phi')} &\qw & \qw & \qw & \EPRur& \
}
\end{split}
\quad .
\end{equation}
All the non-trivial operations are now on the input and output modes, so we commute the displacements through the leftmost beam splitter and pull the circuit taut to get
\begin{equation}\label{displacementpull_QRL}
    \begin{split}
\Qcircuit @C=1.1em @R=1.25em {
		&\lstick{\text{(out)}}&\gate{A(\psi,\phi)}& \gate{D(\mu_{+})}& \qw&\gate{V(\theta_a,\theta_b)}  &  \bsbal{1} &\qw& \rstick{\text{(in)}} \\
		&\lstick{\text{(out)}}&\gate{A(\psi',\phi')}& \gate{D(\mu_{-})}&\bsbal{-1}& \gate{V(\theta_c,\theta_d)} & \qw&\qw& \rstick{\text{(in)}} \\
}
\end{split}
\qquad \; ,
\end{equation}
where the displacement amplitudes are
    \begin{equation}
        \mu_\pm \coloneqq \frac{1}{\sqrt{2}} (\mu_{c,d}\pm\mu_{a,b})
        \, . 
    \end{equation}
This technique can be used to find the two-mode Kraus operators for other macronode lattices including that in Ref.~\cite{larsen2021architecture}; these will be presented in future work. 

\typeout{}
\bibliography{ref}
\end{document}